\documentclass[conference]{IEEEtran}
\IEEEoverridecommandlockouts

\usepackage{soul}
\usepackage{url}

\usepackage{graphicx}
\usepackage{amsmath}
\usepackage{amsthm}
\usepackage{booktabs}
\usepackage{bm}
\usepackage{algorithm}
\usepackage{amsfonts}
\usepackage{algorithmic}
\usepackage{threeparttable}
\usepackage{subfigure}
\usepackage{tabularx, booktabs}

\newcommand{\mb}{\mathbf}

\newcommand{\bs}{\boldsymbol}

\newtheorem{definition}{Definition}
\newcolumntype{P}[1]{>{\centering\arraybackslash}p{#1}}

\def\BibTeX{{\rm B\kern-.05em{\sc i\kern-.025em b}\kern-.08em
    T\kern-.1667em\lower.7ex\hbox{E}\kern-.125emX}}
\begin{document}

\title{Semi-Supervised Variational User Identity Linkage via Noise-Aware Self-Learning}

\author{\IEEEauthorblockN{Chaozhuo Li\IEEEauthorrefmark{1},
		Senzhang Wang\IEEEauthorrefmark{2}, Zheng Liu\IEEEauthorrefmark{1},
		Xing Xie\IEEEauthorrefmark{1}, Lei Chen\IEEEauthorrefmark{3}, 
		Philip S. Yu\IEEEauthorrefmark{4}}
	\IEEEauthorblockA{\IEEEauthorrefmark{1}Microsoft Research Asia,
		\IEEEauthorrefmark{2}Central South University,
		\IEEEauthorrefmark{3}Hong Kong University of Science and Technology,\\
		\IEEEauthorrefmark{4}University of Illinois at Chicago\\
		\{cli, zheng.liu, xingx\}@microsoft.com, szwang@csu.edu.cn, leichen@cse.ust.hk, psyu@uic.edu}}
\maketitle

\maketitle

\begin{abstract}
User identity linkage, which aims to link identities of a natural person across different social platforms, has attracted increasing research interest recently. 
Existing approaches usually first embed the identities as deterministic vectors in a shared latent space, and then learn a classifier based on the available annotations.  However, the formation and characteristics of real-world social platforms are full of uncertainties, which makes these deterministic embedding based methods sub-optimal. In addition, it is intractable to collect sufficient linkage annotations due to the tremendous gaps between different platforms. 
Semi-supervised models utilize the unlabeled data to help capture the intrinsic data distribution, which are more promising in practical usage. However, the existing semi-supervised linkage methods heavily rely on the heuristically defined similarity measurements to incorporate the innate closeness between labeled and unlabeled samples. Such manually designed assumptions may not be consistent with the actual linkage signals and further introduce the noises.  To address the mentioned limitations, in this paper we propose a novel Noise-aware Semi-supervised Variational User Identity Linkage (NSVUIL) model. Specifically, we first propose a novel supervised linkage module to incorporate the available annotations.  Each social identity is represented by a Gaussian distribution in the Wasserstein space to simultaneously preserve the fine-grained social profiles and model the uncertainty of identities. Then, a noise-aware self-learning module is designed to faithfully augment the few available annotations, which is capable of filtering noises from the pseudo-labels generated by the supervised module. The filtered reliable candidates are added into the labeled set to provide enhanced training guidance for the next training iteration. Empirically, we evaluate the NSVUIL model over multiple real-world datasets, and the experimental results demonstrate its superiority.
\end{abstract}


\section{Introduction}

Nowadays, users tend to simultaneously join a variety of social platforms to enjoy different types of services (e.g., LinkedIn for job seeking and Twitter for opinion sharing). 
When a user registers on a social platform, an identity is created to represent her unique personal figure including the demographic information, social relations and published tweets. 
Fusing data from different social platforms contributes to accurately depicting the user figure from multiple perspectives. 
As an indispensable step in cross-platform social mining, user identity linkage (UIL), which aims to link identities of a same natural person across different social networks, has attracted enormous attention considering its significant research challenges and tremendous practical values. 
A myriad of real-life applications can take advantage of successful UIL, such as social recommendation \cite{shu2017user}, information diffusing prediction \cite{zafarani2014users,wang2014mmrate,zhan2015influence} and network dynamics analysis \cite{wang2015burst,zafarani2016users}.

%
%
%
%
%
%
%
%
%
%
%
%
%

Most existing UIL approaches are supervised models, which need a large number of annotations to train a classifier or ranker to separate linked identity pairs from unlinked ones  \cite{motoyama2009seek,vosecky2009user,iofciu2011identifying,perito2011unique,peled2013entity,zhang2014online,mu2016user,man2016predict,nie2016identifying}. 
Considering the boundaries between different platforms, it is extremely expensive and time-consuming to manually collect sufficient annotations. 
In order to eliminate the dependence on annotations, unsupervised approaches are proposed to link identities according to some pre-defined discriminative features (e.g., screen name) \cite{labitzke2011your,liu2013s,lacoste2013sigma,riederer2016linking} or the distribution similarities of social spaces \cite{li2018distribution}. 
However, unsupervised methods often suffer from high false rate as no annotation is incorporated as the learning guidance \cite{Li2019}. 
Semi-supervised learning can utilize both labeled and unlabeled data to help capture the shape of the intrinsic data distribution, which is more promising  in practice. 
Existing semi-supervised UIL approaches usually first capture the inherent similarities between the labeled identities and unlabeled ones, and then incorporate these unsupervised information as complementary to facilitate the identity linkage. 
Various types of similarity measurements (e.g., topological connection \cite{korula2014efficient,man2016predict}, text similarity \cite{zhang2015cosnet,zhong2018colink} and distribution closeness \cite{Li2019,li2019partially}) have been extensively studied by previous works. 

However, such heuristically defined similarities may be inconsistent with the actual linkage annotations. 
For example, a same natural person may publish work-related news in LinkedIn while sharing the daily life in Twitter, leading to the textual dissimilar.   
Generally, annotations can provide more reliable and task-relevant learning signals compared with unsupervised information.  
If such supervised signals can be automatically and accurately augmented, the linkage performance is expected to be further boosted. 
Although this self-learning strategy seems to be a promising solution, only a few UIL works focus on this topic \cite{nguyen2020structural}. 
The major reason lies in that linking social identities across different platforms is quite challenging due to the enormous search space (the size of $m \times n$ given two social networks with $m$ and $n$ identities respectively). 
The candidates with high confidence generated by supervised linkage models may still have a large chance to be incorrect. 
As shown in Table \ref{tab:examples_of_noises}, we select top 50 unlabeled pairs with the highest confidence scores (i.e., linkage probabilities) and evaluate whether they are correctly matched.  From the results of two SOTA methods (CoLink and SNNA$_{o}$), one can clearly see that nearly half of these top samples are bad cases.   
Vanilla self-learning models directly add these untrustworthy pseudo-labels into the ground truth set, which may introduce and enlarge noises because early mistakes could reinforce themselves in the iteratively learning process \cite{levatic2017self}. 
Hence, a noise-aware self-learning mechanism designed to filter noises from the confident candidates is indispensable.    

\begin{table}
	\centering
	\begin{threeparttable}
		\caption{An example of the accuracy scores of the top confident pairs learned by different linkage models over two datasets.}
		\begin{tabular}{P{1.6cm}P{2.2cm}P{2.2cm}P{0.7cm}}
			\toprule
			\multicolumn{1}{c}{Methods}&\multicolumn{1}{c}{Twitter-Flickr}&\multicolumn{1}{c}{Weibo-Douban}\cr
			\midrule
			CoLink&0.46&0.48\cr
			SNNA$_{o}$&0.56&0.52\cr
			\midrule
			NSVUIL$_{sup}$&0.66&0.69\cr
			NSVUIL$_{sl}$&0.88&0.87\cr
			\bottomrule
		\end{tabular}
		\label{tab:examples_of_noises}
		\vspace{-4mm}
	\end{threeparttable}
\end{table}

In addition, high-quality user representations also contribute to improving the quality of pseudo-labels.   
The motivation is that in the latent representation space, social identities with similar profiles (e.g., relations and posted microblogs) as well as the linkage labels should be closely distributed and can be easily clustered into the same group. 
Then it would be much easier to learn a desirable linkage model to generate reliable pseudo-labels.  
Existing approaches usually learn user representations by simple and naive models (e.g., bag-of-words vectors for the textual data  \cite{li2018distribution,Li2019} and matrix factorization for the social relation encoding \cite{zhang2014online}), which cannot comprehensively reflect the user characteristics in terms of social relations and posted texts. 
Another limitation of existing methods is that each user is embedded as a deterministic point in the representation space, which ignores the uncertainties in social networks. 
Human behavior in the social platforms is multi-faceted which makes the generation of social relations uncertain \cite{zhu2018deep}. 
It is widely recognized that for the identities with few social relations, their representations bear more uncertainties than others due to the sparse user generated data \cite{zhu2018deep,jiang2020convolutional}. 
Without considering the uncertainty of social networks, the learned user embeddings will be less effective in the UIL task. 
Thus, we also focus on how to learn desirable user embeddings to improve the linkage performance. 

%
%

In this paper, we propose a deep learning based framework NSVUIL for \textbf{N}oise-aware \textbf{S}emi-supervised \textbf{V}ariational \textbf{U}ser  \textbf{I}dentity \textbf{L}inkage task.   
Our motivation lies in jointly learning high quality user representations for supervised identity linkage, and identifying noisy pairs in the pseudo-labels for unsupervised self-learning.  
Specifically, we first propose a novel supervised identity linkage module, which consists of the hierarchical attention-based identity modeling component and the variational identity linkage component. 
Hierarchical identity modeling introduces the attention mechanism to fuse the fine-grained semantic information with social relations, which is able to learn desirable user representations by capturing the task-relevant representative user information.   
Variational identity linkage component transforms the user representation to a Gaussian distribution $\mathcal{N}(\bs{\mu}, \bs{\Sigma})$, where the mean vector $\bs{\mu}$ implies the representative features of the users, and the variance vector $\bs{\Sigma}$ innately represents the uncertainty. 
Compared with the deterministic embeddings used in the existing works, Gaussian distribution based representations are capable of simultaneously preserving the informative user features and capturing the uncertainties of social networks.   
After that, different from the vanilla self-learning strategies, a generative noise-aware unsupervised module is introduced to filter the noise from pseudo-labels generated by the supervised linkage module. 
Then, the rest reliable samples are added into the ground truth set to provide enhanced training guidance. 
These two modules are iteratively trained under a unified learning framework. 
The proposed NSVUIL model is thoroughly evaluated on five pairs of real-life datasets, and the experimental results demonstrate its superiority.  

We summarize our main contributions as follows: 

\begin{itemize}
	\item To the best of our knowledge, we are the first to study the task of noise-aware variational social identity linkage. 
	The proposed NSVUIL model is capable of effectively filtering noises  and learning a desirable linkage model simultaneously.      
	\item We propose a hierarchical-attention based variational linkage module to capture the cross-platform supervised signals, and a novel latent variable based generative module to perform unsupervised noise-aware self-learning.   
	\item Extensively, we evaluate NSVUIL on five groups of datasets. Experimental results demonstrate the superior performance of the proposed approach.
	
\end{itemize}

%

\begin{figure*}
	\centering
	\includegraphics[width=0.90\textwidth]{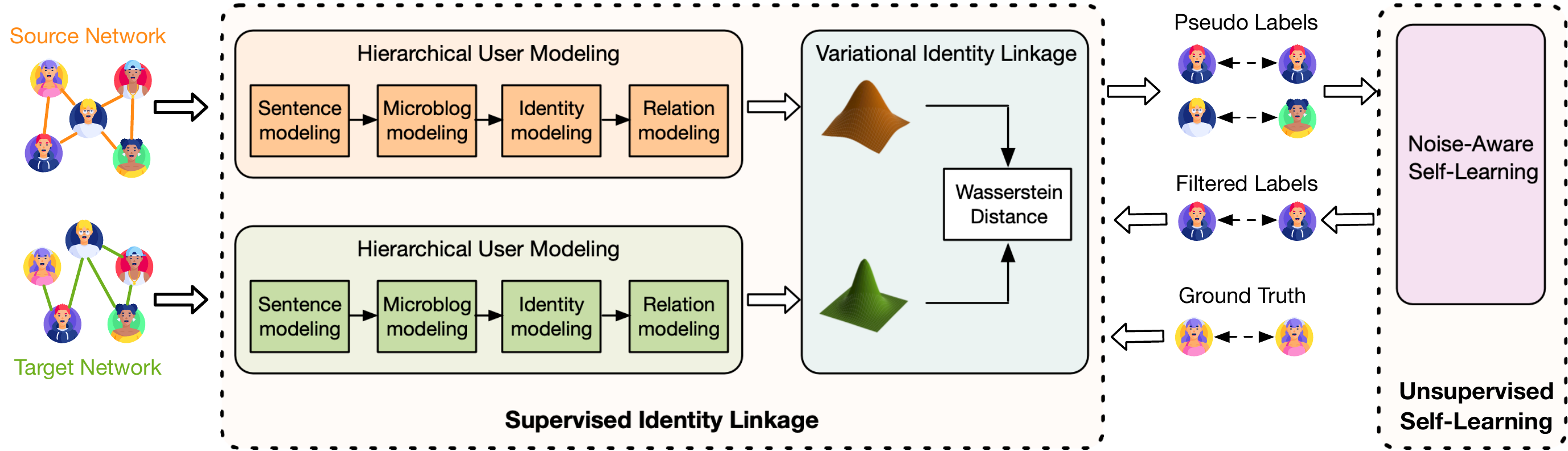}
	\caption{Framework of the proposed NSVUIL model.}
	\label{fig:framework} 
	\vspace{-4mm}
\end{figure*}

\section{Related Work}
In this section, we will summarize and introduce the related works.  
Existing user identity linkage approaches can be roughly categorized into supervised, semi-supervised and unsupervised methods. 
Most existing methods are supervised, which view the studied task as a ranking or classification problem to locate the candidates (identity pairs) with the highest linkage probabilities \cite{vosecky2009user,motoyama2009seek,iofciu2011identifying,perito2011unique,peled2013entity,zhang2014online,mu2016user,man2016predict,nie2016identifying}. 
Man et al. \cite{man2016predict} keep major structural regularities of networks by leveraging the observed anchor links as supervised information. After that, a stable cross-network mapping is learned to link identical identities.
Mu et al. \cite{mu2016user} optimize objective function jointly with matching/non-matching pairs and intra-platform relation constraints across different platforms. 
Liu et al. \cite{liu2014hydra} propose to model heterogeneous behavior by long-term behavior distribution analysis and multi-resolution temporal information matching. 
Zhang et al. \cite{zhang2018mego2vec} introduce the ego graphs to model the influence of neighbors and further propose the linkage model on the ego-graph level. 
Zhang et al. \cite{zhang2014online} utilize the domain-specific prior knowledge as guidance and a probabilistic classifier is applied on a set of extensive profiles extracted for social network profile linkage. 
Zhang et al. \cite{zhang2019graph} introduce the popular graph convolutional network and jointly capture local and global information for UIL.

Considering the  tremendous information gaps between different social networks, it is labor costing and time-consuming to achieve sufficient annotations. 
Thus, several unsupervised approaches are proposed to automatically locate the linked identity pairs \cite{labitzke2011your,liu2013s,lacoste2013sigma,riederer2016linking,li2018distribution}.  
Liu et al. \cite{liu2013s} propose to first automatically generate a set of training samples according to the rareness of the user names in two social networks, and then use these samples to train a binary classifier.
Lacoste-Julien et al. \cite{lacoste2013sigma} propose a greedy approach to align the attributes of users according to the heuristic text similarities.
POIS \cite{riederer2016linking} uses the trajectory-based attribute features to link user identities. 
Li et al. \cite{li2018distribution} consider all the users in a social network as a whole and perform user identity linkage from the level of user space distribution. 
Earth mover’s distance (EMD) is introduced as the measure
of distribution closeness to learn the  projection function. 
Although unsupervised approaches eliminate the reliance on the annotations, they often suffer from low performance as no annotation is incorporated to provide the learning guidance.

Recently several semi-supervised methods are proposed to incorporate  unlabeled data to capture the inner data distribution, which are  more  promising  to perform user identity linkage task.
Korula et al. \cite{korula2014efficient} design a simple, local, and efficient algorithm with provable guarantees and utilize a small fraction of account linkage to  identify a very large fraction of the network.
By considering both local and global consistency, Zhang et al. \cite{zhang2015cosnet} propose an energy-based model to link user
identities and develop an efficient subgradient algorithm to convert the original energy-based objective function into its dual form.
Zhong et al. \cite{zhong2018colink} propose one attribute-based model and one relationship-based model and make them reinforce each other iteratively in a co-training framework for identity linkage.
Li et al. \cite{Li2019} incorporate the isomorphism across social networks as complementary to link identities from the distribution level. Three adversarial learning based models are proposed to minimize the Wasserstein distance between two social distributions. 
They further propose a multi-platform based social identity linkage model with partially shared generators and discriminators, which is also defined under the adversarial learning framework \cite{li2019partially}. 
Different from existing approaches, in this paper we aim to link identities in the self-learning manner. 
Existing self-learning methods usually directly add the pseudo-labels into the training labels, which may introduce and enlarge noises \cite{nguyen2019self}. 
Our motivation lies in jointly learning desirable user representations for supervised identity linkage, and identifying noisy samples in the selected confident unlabeled pairs for unsupervised self-learning.  

\section{Problem Definition}

A social network is formally defined as $\mb{N} =  \{\mb{U}, \mb{E}\}$, in which $\mb{U} = \{u_{1}$, $u_{2}$, $\cdots$, $u_{n}\}$ denotes the user identities in the social network. 
Matrix $\mb{E} \in \{0,1\}^{n \times n}$ is the adjacency matrix of the undirected social relation graph. 
Each user is associated with a set of posted microblogs and user-defined demographic information (e.g., age, gender, etc.).  
The studied problem is formally defined as follows:  
\begin{definition}
	\emph{\textbf{Semi-supervised User Identity Linkage.} 	
		The input two social networks are formally defined as the source network $\mb{S} = \{\mb{U_{S}}, \mb{E_{S}}\}$ and the target one $\mb{T} =  \{\mb{U_{T}}, \mb{E_{T}}\}$. 
		In addition, a few available linked identity pairs are defined as $\mb{A} =\{(u_{s}, u_{t}) | u_{s} \in \mb{U_{S}}, u_{t} \in \mb{U_{T}} \}$, in which two identities $u_{s}$ and  $u_{t}$ belong to the same person.  
		We aim to learn a matching function $f$ to locate the rest aligned identity pairs $\mb{Y} =\{(u_{s}, u_{t}) | u_{s} \in \mb{U_{S}}, u_{t} \in \mb{U_{T}}, (u_{s}, u_{t}) \notin \mb{A} \} $.  
	}
\end{definition}
Following previous works \cite{mu2016user,zhang2019graph}, two input networks are partially aligned, which means a set of identities in one social network do not have the matching ones in the other network.

\section{Methodology}
\subsection{Framework}
The framework of NSVUIL is shown in Fig. \ref{fig:framework}.  
NSVUIL includes two major modules: the supervised identity linkage module to capture the cross-platform linkage signals and the unsupervised self-learning module to augment the annotations. 
Two social networks $\mb{S}$ and $\mb{T}$ first go through the supervised identity linkage module as shown in the center part of Fig. \ref{fig:framework}.  
The hierarchical user modeling is proposed to capture the fine-grained social information in terms of sentences, microblogs and social relations. 
The attention mechanism is introduced to select the task-relevant informative data. 
Then in the variational identity linkage part, considering the Gaussian distribution innately represents the uncertainty property \cite{vilnis2014word}, it is promising to represent an identity by Gaussian distributions, i.e. the mean and the variance, rather than a point vector to incorporate the uncertainty. 
The user representations are transformed into the Gaussian distributions, and a Wasserstein distance based loss is proposed to minimize the distribution distances between  linked identities in the ground truth set.           
After fully training the supervised model, we select the most confident unlabeled samples (i.e., pseudo-labels), and feed them into a noise-aware self-learning module to further filter noises as shown in the right part of Fig. \ref{fig:framework}. 
Finally, the filtered reliable annotations are added into the ground truth set for the next iteration training. 

\begin{figure}
	\centering
	\includegraphics[width=0.42\textwidth]{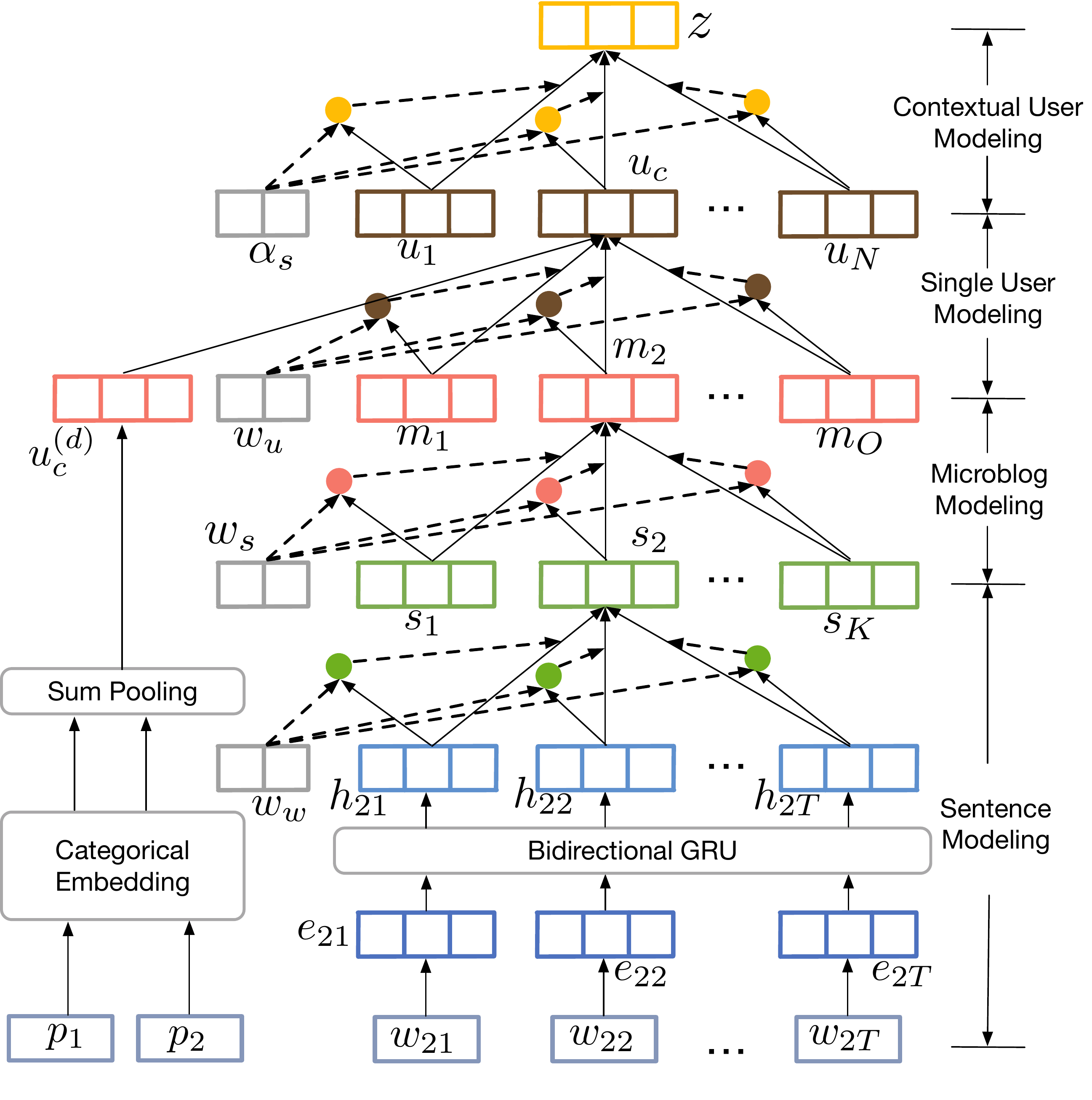}
	\caption{An illustration of the hierarchical user modeling part. Circles indicate the attention in each layer. From top to bottom, $u_{c}$ denotes the input user and other $u_{i}$s are her neighbors. Assuming the input user has published $O$ microblogs, we select $m_{2}$ as an example to demonstrate the following modeling parts. The lower layers are presented in the similar way.}
	\label{fig:usermodeling} 
	\vspace{-4mm}
\end{figure}
   
\subsection{Supervised User Identity Linkage}

In this subsection, the details of two major components (hierarchical user modeling and variational identity linkage) in the supervised identity linkage module are presented.  
\subsubsection{Hierarchical User Modeling}
Existing approaches usually focus on how to design an effective linkage function but ignore the significance of user modeling part. 
For example,  \cite{li2018distribution,li2019partially} employ the bag-of-words method to represent the posted microblogs, which assumes all words are independent of each other but ignores the latent semantic correlations among different words. 
Some methods \cite{lim2015mytweet,zafarani2015user} only incorporate the user attributes but discard the important social relations.  
Meanwhile, the microblogs and user attributes in the social networks are usually missing, fake or meaningless.  
It is intractable to select crucial and task-relevant information from such chaos data.  
Here we propose a hierarchical attention-based user modeling component to effectively capture the fine-grained and complete social user information.

\noindent \textbf{Sentence Modeling} Existing approaches usually view a posted microblog as a single sentence \cite{lim2015mytweet,zafarani2015user}. However, the maximum length of the users can post in a single microblog is expanded from 140 characters to 280 characters in the Twitter platform, which means a single microblog may consist of several sentences. 
Hence, we aim to learn the desirable  social user representations starting from the fine-grained sentence level. 
As shown in Figure \ref{fig:usermodeling}, given a sentence $s_i$ containing $T$ words $[w_{i1}, w_{i2}, \cdots, w_{iT}]$, it is converted into a sequence of $d$-dimensional vectors $[\mb{e_{i1}}, \mb{e_{i2}}, \cdots, \mb{e_{iT}}]$ with a pre-learned word embedding matrix. 
Then, a bidirectional GRU \cite{cho2014learning} is introduced to learn contextual word representations $\mb{h_{it}}$, which contains a forward GRU reading the sentence from $w_{i1}$ to $w_{iT}$ and a backward GRU reading words in the opposite direction. 
The contextual representation $\mb{h_{it}}$ of word $w_{it}$ is obtained by concatenating the forward hidden state and the backward state of  $w_{it}$ learned from two GRUs. 
After that, the attention mechanism is employed to learn the semantic importance of words. 
The insight is that different words in the same sentence may have different informativeness. 
For example, in the sentence ``\textit{The apple I bought today is yummy}'', the word ``\textit{yummy}'' is more representative than the word ``\textit{today}'' on the depiction of the entity ``\textit{apple}''. 
The attention score of the $t$-th word in the sentence $s_{i}$ is formalized as follows:
\begin{equation}
	\label{word_attention}
	\begin{aligned}
		a_{it}^{(w)} &= \sigma(\mb{w_{w}} \cdot \mb{h_{it}} + b_{w}), \\
		\alpha_{it}^{(w)} &= \frac{\exp(a_{it}^{(w)})}{\sum_{j=1}^{T}\exp(a_{ij}^{(w)})}
	\end{aligned}
\end{equation}
in which $\mb{w_{w}} \in \mathbb{R}^{d_{w}}$ and $b_{w} \in \mathbb{R}$ are the parameters in the attention network. $\sigma$ is the activation function. $\alpha_{it}$ denotes the normalized importance of the $t$-th word compared with other words. The final representation of the sentence $s_{i}$ is the weighted sum of the word representations based on the learned attention weights: 
\begin{equation}
	\begin{aligned}
		\mb{s_{i}} = \sum_{t=1}^{T} \alpha_{it}^{(w)} \mb{h_{it}}
	\end{aligned}
\end{equation}

%
%

\noindent \textbf{Microblog Modeling} Microblog modeling aims to aggregate the learned sentence representations into a final representation of the belonged microblog. 
As shown in Figure \ref{fig:usermodeling}, a microblog $m_{i}$ consists of $K$ sentences $[s_{i1}, s_{i2}, \cdots, s_{iK}]$. 
Considering the number of sentences in a microblog is comparatively small, here we directly employ attention strategy without the contextual representation learning part (GRUs). 
Similar to Formula (\ref{word_attention}),  we use a sentence-level attention network to help our model select and attend to important sentences: 
\begin{equation}
	\begin{aligned}
		a_{ik}^{(s)} &= \sigma(\mb{w_{s}} \cdot \mb{s_{ik}} + b_{s}), \\
		\alpha_{ik}^{(s)} &= \frac{\exp(a_{ik}^{(s)})}{\sum_{j=1}^{K}\exp(a_{ij}^{(s)})}, \\
		\mb{m_{i}} &= \sum_{k=1}^{K} \alpha_{ik}^{(s)} \mb{s_{ik}}
	\end{aligned}
\end{equation}
in which $\mb{w_{s}}$ and $b_{s}$ are the parameters in the sentence-level attention network. $\alpha_{ik}^{(s)}$ indicates the relative importance of the $k$-th sentence. The final representation of the microblog $\mb{m_{i}}$ is the summation of the  sentence representations weighted by the learned attention
scores.

\noindent \textbf{Single User Modeling} 
Assuming user $u_{i}$ has published $O$ microblogs $[m_{i1}, m_{i2}, \cdots , m_{iO}]$, here we aim to learn the user representation by weighted aggregating these microblogs.
For example, the microblog ``\textit{Knicks Rules!}'' reflects more user preferences than ``\textit{It's raining outside}''.  
Thus, we further employ a microblog-level attention network to distinguish informative microblogs from less informative ones.  
Considering the microblog-level attention network is similar to the ones used in the sentence and microblog modeling, here we skip the detailed calculations.    
We denote the learned user representation from the posted microblogs as $\mb{u_{i}^{(t)}}$. 
Users also have some demographic information such as interest tags, gender and locations, which also help depict user figures. 
In order to highlight these demographic information, we view them as the categorical features and represent them with the embedding matrix $\mb{P_{d}} \in \mathbb{R}^{V_{d} \times d_{d}}$, where $V_{d}$ is the demographic feature set and $d_{d}$ is the embedding size. $\mb{P}_{d}$ is randomly initialized and is trainable during the learning process. 
Considering users may have different numbers of demographic features, here we employ  max-pooling  to learn a single  demographic embedding $\mb{u_{i}^{(d)}}$.
The final user representation $\mb{u_{i}}$ is the concatenation of user representation $\mb{u_{i}^{(t)}}$ learned from microblogs and the demographic embedding $\mb{u_{i}^{(d)}}$.

\noindent \textbf{Contextual User Modeling}
Social relations have been proven to be pivotal in the identity linkage task \cite{li2018distribution,liu2016aligning}.  
Following the popular graph neural networks \cite{velivckovic2017graph}, here we propose to encode the social relationships into the contextual user representations by weighted aggregating the social neighbors.  
Fusing complementary information from  neighbors also contributes to alleviating the data sparsity issue as the user attributes are self-defined and can be missing or sparse. 

Given a center user $u_{c}$ and her neighbors $[u_{c1}, u_{c2}, \cdots, u_{cN}]$, we also employ attention strategy to learn the neighbor weights properly and then weighted aggregate the contextual semantic information.  
The attention score between the center user $u_{c}$ and her neighbor $u_{cn}$ is calculated as follows: 
\begin{equation}
	\alpha_{cn} = \frac{\exp (\sigma(\mb{a_{s}} \cdot [\mb{u_{c}}||\mb{u_{cn}}]))}{\sum_{i = 1}^{N} \exp(\sigma(\mb{a_s} \cdot [\mb{u_c}||\mb{u_{ci}}]))}
\end{equation}
in which $||$ is the concatenation operation and $\mb{a_{s}}$ is the local-level attention vector for the neighbors. 
Learned attention score $\alpha_{cn}$ denotes how important neighbor $u_{cn}$ will be for the center node $u_{c}$.
Then the relation-enhanced embedding $\mb{z_{c}^{(r)}}$ can be aggregated by the neighbor’s features with the corresponding
coefficients as follows:
\begin{equation}
	\mb{z_{c}^{(r)}} = \sigma(\sum_{i = 1}^{N} \alpha_{ci}\cdot \mb{u_{ci})}  
\end{equation}
Finally, the single user representation $u_{c}$ and the contextual embedding $z_{c}$ are concatenated as the final user embedding:   
\begin{equation}
	\begin{aligned}
		\mb{z} = [\mb{z_{c}^{(r)}}, \mb{u_{c}}]
	\end{aligned}
\end{equation} 

\subsubsection{Variational Identity Linkage}
\begin{figure}
	\centering
	\includegraphics[width=0.46\textwidth]{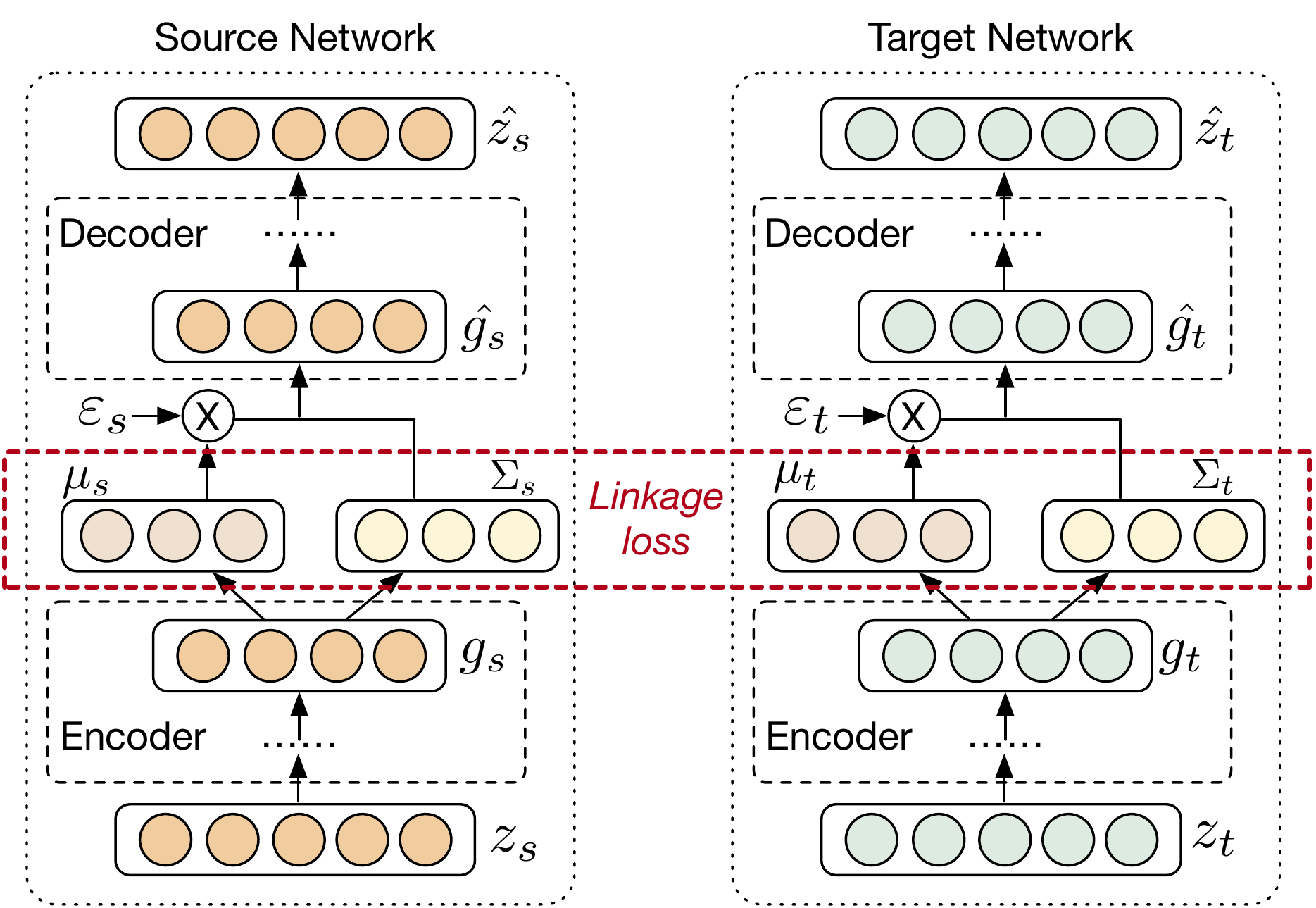}
	\caption{An illustration of the variational linkage module.}
	\label{fig:supervised} 
	\vspace{-4mm}
\end{figure}


Given the learned user representation $\mb{z}$, most existing UIL methods directly leverage these deterministic vectors to learn a classifier or ranker as the linkage model. 
However, as discussed in the introduction section, the formation and evolution of the social networks are full of uncertainties. 
Meanwhile, Gaussian distributions innately represent the uncertainty property. 
Therefore, we propose to represent users by Gaussian distributions, i.e., the means and the variances, rather than the point vectors to incorporate the uncertainty. 
Each user $u_{i}$ is represented by a lower-dimensional Gaussian distribution $\mathcal{N}(\bs{\mu_{i}}, \bs{\Sigma_{i}})$, in which $\bs{\mu_{i}}$ denotes the position of user in the embedding space and $\bs{\Sigma_{i}}$ investigates the uncertainty of the user. 
Following the popular variational autoencoder \cite{kingma2013auto}, the variational identity linkage module includes the following three components as shown in Figure \ref{fig:supervised}. 

\noindent \textbf{Encoder}  
Encoder aims to project the user representations from different social networks into a shared latent space. 
The user representation $\mb{z}$ learned by the hierarchical user modeling is fed into the encoder to provide comprehensive user profiles. 
Here we take the source identity  as an example.  
Encoder is implemented as a two-layer multi-layer perceptron:      
\begin{equation}
	\begin{aligned}
		\mb{g_{s}}^{(1)} &= \sigma(\mb{W^{(1)}} \times \mb{z_{s}} + \mb{b}^{(1)}) \\
		\mb{g_{s}} &= \sigma(\mb{W^{(2)}} \times \mb{g_{s}}^{(1)} + \mb{b}^{(2)})
	\end{aligned}
\end{equation} 
in which $\mb{W^{(i)}}$ and $\mb{b^{(i)}}$ are trainable parameters. 

\noindent \textbf{Variational Layer} 
The variational layer projects the latent deterministic embeddings to the Gaussian space to preserve the uncertainties and informative user features.  
Two one-layer perceptron networks are introduced to learn the Gaussian embeddings by estimating the mean and variance vectors, respectively:   
\begin{equation}
	\begin{aligned}
		\bs{\mu_{s}} &= \sigma(\mb{W_{mu}} \times \mb{z_{s}} + \mb{b_{mu}}) \\
		\bs{\Sigma_{s}} &= \sigma(\mb{W_{var}} \times \mb{z_{s}} + \mb{b_{var}}) \\
	\end{aligned}
\end{equation} 
in which $\mb{W_{mu}}$, $\mb{W_{var}}$, $\mb{b_{mu}}$ and $\mb{b_{var}}$ are trainable parameters. 
Each user is represented by a Gaussian distribution, in which the mean vector $\bs{\mu_{s}}$ finds an approximate
position of the identity and the variance term $\bs{\Sigma_{s}}$ captures the uncertainty.

\noindent \textbf{Decoder} 
Decoder aims to recover the original user representations from the latent Gaussian distributions, which guarantees the informative features of social users are properly preserved. 
One inherent challenge is that we need to sample the latent user embedding from her associated Gaussian distribution. 
However, this sampling process is a non-continuous operation and has no gradient, leading to the failure of the back-propagation. 
Inspired by the variational autoencoders \cite{kingma2013auto}, we introduce the ``reparameterization trick'' to tackle this challenge. 
A noise vector $\bs{\epsilon_{s}}$ is sampled from a standard Gaussian distribution. 
Then the sampled user embedding can be achieved by the combination of the noise vector and the corresponding Gaussian distribution: 
\begin{equation}
	\begin{aligned}
		\mb{\hat{g}_{s}} = \bs{\mu_{s}} + (\bs{\Sigma_{s}})^{1/2} \circ \bs{\epsilon_{s}}
	\end{aligned}
\end{equation} 
in which $\circ$ denotes the element-wise multiplication. 
In this way, the gradients can be back-propagated through the entire model. 
Decoder also contains a two-layer multi-layer perceptron to map the sampled embeddings back to the input space, which is the same to the one used in the encoder.

\subsubsection{Objective Function}
\label{onjective}
The objective function of the supervised user identity linkage includes two types of losses: the identity linkage loss and the reconstruction loss. 
In this subsection, we will present the details of these losses. 

\noindent \textbf{Identity Linkage Loss}  
Intuitively, the identity linkage loss ensures the matched identities should be closer than the unmatched ones. 
Recall that two Gaussian distributions of source identity and target identity,  $\mathcal{N}(\bs{\mu_{s}}, \bs{\Sigma_{s}})$  and  $\mathcal{N}(\bs{\mu_{t}}, \bs{\Sigma_{t}})$, are estimated by the corresponding variational user modeling parts. 
An appropriate statistical measurement is required to measure the distance between these two Gaussian distributions. 
Kullback-Leibler (KL) divergence and the $p$-th Wasserstein distance are two popular distribution distances, in which Wasserstein distance is more promising as KL divergence is not symmetric and does not satisfy the triangle inequality \cite{zheng2019deep}. 
Moreover,  if two distributions are non-overlapping, the Wasserstein distance can still measure the distance between them, while KL-divergence fails and leads to vanishing gradients. 
Thus, we adopt the $p$-th Wasserstein distance as the measurement. 
The calculation of the general-formed Wasserstein distance is limited by the heavy computational cost. 
Considering the identities are represented by Gaussian distributions, we explore the 2-th Wasserstein distance as it has closed form solution to speed-up the calculation and reduce the computational costs. 

Given the input source identity $u_{s}$, we first achieve its matched target identity $u_{t}$ and one unmatched target identity $\hat{u}_{t}$ by negative sampling.  
The Wasserstein-based identity linkage loss is proposed as:
\begin{equation}
	\label{linkageloss}
	\begin{aligned}
		\mathcal{L}_{l} = &- \sum_{(u_{s}, u_{t}, \hat{u}_{t})} \ln \sigma \{ W_{2}(\mathcal{N}(\bs{\mu_{s}}, \bs{\Sigma_{s}}), \mathcal{N}(\bs{\mu_{\hat{t}}}, \bs{\Sigma_{\hat{t}}}))  \\
		&- W_{2}(\mathcal{N}(\bs{\mu_{s}}, \bs{\Sigma_{s}}), \mathcal{N}(\bs{\mu_{t}}, \bs{\Sigma_{t}}) )\} + \lambda (\parallel\bs{\Omega}_{s}\parallel^2_{2} + \parallel\bs{\Omega}_{t}\parallel^2_{2})
	\end{aligned}
\end{equation}  
in which $\sigma$ denotes the sigmoid function. $\bs{\Omega_{s}}$ and $\bs{\Omega_{t}}$ are the trainable parameters. Hyper-parameter $\lambda$ represents
the weight on the regularization terms. 
The $W_{2}$ Wasserstein distance has a closed solution as follows \cite{givens1984class}: 
\begin{equation}
	\begin{aligned}
		W_{2}(\mathcal{N}(\bs{\mu_{s}}, \bs{\Sigma_{s}}), \mathcal{N}(\bs{\mu_{t}}, \bs{\Sigma_{t}})) = &\parallel\bs{\mu_{s}}-\bs{\mu_{t}}\parallel^2_{2}  \\ 
		+ Tr (\bs{\Sigma_{s}} + \bs{\Sigma_{t}} - &2 * (\bs{\Sigma_{s}}^{1/2}\bs{\Sigma_{t}}\bs{\Sigma_{s}}^{1/2})^{1/2})
	\end{aligned}
\end{equation}  
Loss (\ref{linkageloss}) seeks to maximize the Wasserstein distance of a negative pair $(u_{s}, \hat{u_{t}})$ and minimize the distance of a
positive pair $(u_{s}, u_{t})$. 

\begin{algorithm}[tb]
	\caption{Training process of the supervised linkage part}
	\label{alg:algorithm1} 
	\begin{flushleft}
		\textbf{Require}: source network $S$, target network $T$ and the annotation set $A$  
	\end{flushleft}
	\begin{algorithmic}[1] 
		\STATE Initialize parameters in $\Omega_{t}$ and $\Omega_{s}$  
		\WHILE{$\mathcal{L}$ does not converage}
		\STATE Sample a batch  of linked pairs $b_{a}$ from set $A$ \\
		\STATE Sample a batch of unlinked pairs $b_{u}$ from $U_S \times U_T$\\
		\STATE Combine $b_{a}$ and $b_{u}$ as the training batch
		\STATE In-batch negative sampling for Formula (\ref{linkageloss}) \\
		\STATE Calculate loss following Formula (\ref{finalloss}) and update parameters    \\
		\ENDWHILE
	\end{algorithmic}
\end{algorithm}

\noindent \textbf{Reconstruction Loss}  
In order to ensure the informative features of social users are appropriately preserved in the Gaussian distributions, the user representation $\bs{z_{s}}$ from the source network is expected to be reconstructed from the distribution $\mathcal{N}(\bs{\mu_{s}}, \bs{\Sigma_{s}}$). 
The reconstruction loss is defined as: 
\begin{equation}
	\begin{aligned}
		\mathcal{L}_{r}^{(s)} = \sum_{s \in \mb{U_{s}}}  \parallel\mb{z_{s}} - \mb{\hat{z}_{s}}\parallel^{2} + \frac{1}{2} (-\log \bs{\Sigma_{s}} + (\bs{\mu_{s}})^{2} + \bs{\Sigma_{s}} - 1)
	\end{aligned}
\end{equation}  
in which $\mb{U_{s}}$ is the identity set in the source network.  
The first term models the reconstruction capacity.  
However, only the first term will force the variables to be stable during the training process, i.e., the variance of each variable embedding will tend to be zero.  
Therefore, following previous works, the regularization term is added to force the posterior distribution to be close to the prior one.

The final objective function of the supervised identity linkage module is the weighted combination of linkage loss and the reconstruction losses: 
\begin{equation}
	\label{finalloss}
	\begin{aligned}
		\mathcal{L} = \mathcal{L}_{l} + \beta (\mathcal{L}_{r}^{(s)} + \mathcal{L}_{r}^{(t)})
	\end{aligned}
\end{equation} 
in which $\beta$ is a hyper-parameter to balance the importance of these two losses. 
The training procedure of the supervised identity linkage module is shown in Algorithm \ref{alg:algorithm1}. 
$\bs{\Omega_{s}}$ and $\bs{\Omega_{t}}$ are parameter sets of user modeling modules in social network and target network, respectively. 
The unlinked pairs are also incorporated into the training batches as shown in line 5, which strengthens the power of reconstruction loss and enlarges the candidate space of negative sampling.   
Parameters are updated until the model converges.

\subsection{Noise-Aware Self-Learning Module}

The learned supervised linkage module can locate confident unlabeled identity pairs with the highest linkage probabilities as pseudo-labels.  
Following the identity linkage loss as shown in Formula (\ref{linkageloss}), the unlabeled identity pairs with the shortest Wasserstein distances are selected as pseudo-labels. 
Existing self-learning methods directly add these pairs into the training set to start the next training iteration. 
Considering the search space is quite huge ($|\mb{U_{S}}| \times |\mb{U_{T}}|$), the high confident pairs may not be completely reliable and contain many noises as shown in Table \ref{tab:examples_of_noises}. 
Directly adding these noisy pseudo-labels into the labeled training set may introduce and enlarge noises because early mistakes could reinforce themselves in the iteratively learning process, leading to undesirable performance \cite{nguyen2019self}.

Here we introduce a generative latent variable-based component to filter noisy pairs \cite{lubin2019aligning}. 
Recall that each user $u_i$ is represented by a Gaussian distribution. 
In this subsection, we focus on identifying noises from the pseudo-labels, in which the representative user features are more important than the  uncertainty.  
Hence, we only use the informative mean vectors $\bs{\mu_{i}}$ to represent the identities. 
Given the mean vector $\bs{\mu_{s}}$ of a source identity,  the mean vector of its matched target identity  $\bs{\mu_{t}}$ is generated by two steps. 
Considering the pseudo-labels are either correct samples or noises, a latent distribution taking one of two possible values should be defined, on which the Bernoulli distribution thrives. 
We first sample a random variable $q$ from a Bernoulli distribution with probability $\gamma$:
\begin{align}
	q \sim \text{Bernoulli}(\gamma)
\end{align}
in which $\gamma$ is a trainable parameter. 
$q = 1$  means the sampled $\bs{\mu_{t}}$ and $\bs{\mu_{s}}$ are matched, and then $\bs{\mu_{t}}$ should be sampled from the following distribution:
\begin{align}
	\bs{\mu_{t}} \sim \mathcal{N}(\bs{\mu_{s}}, \bs{\sigma_{p}}^2I), \ \ \ \ \ q=1 \ \text{(`matched')}
\end{align} 
Otherwise $\bs{\mu_{t}}$ is sampled from:
\begin{align}
	\bs{\mu_{t}} \sim \mathcal{N}(\bs{\hat{\mu}}, \bs{\hat{\sigma}}^2I), \ \ \ \ \ q=0 \, \text{(`noise')}
\end{align} 
in which $\mathcal{N}$ is the Gaussian distribution, $\bs{\hat{\mu}}$ and $\bs{\hat{\sigma}}$ are the mean and variance vectors of the noise distribution,  and $\bs{\sigma_{p}}$ denotes the variance of the positive distribution.  All means and variances are trainable parameters. 
Then, the conditional probability of $\bs{\mu_{t}}$ is formatted as  the combination of two Gaussian distributions:
$$f(\bs{\mu_{t}}|\bs{\mu_{s}}) = (1 \!-\! \gamma) \mathcal{N}(\bs{\hat{\mu}}, \bs{\hat{\sigma}}^2I) + \gamma  \mathcal{N}(\bs{\mu_{s}}, \bs{\sigma_{p}}^2I).$$ 
The objective is to maximize the likelihood function:  
\begin{align}
	\mathcal{L}_{sl}(\gamma, \bs{\sigma_{p}}, \bs{\hat{\mu}}, \bs{\hat{\sigma}}) =  \sum \log  f(\bs{\mu_{t}}|\bs{\mu_{s}})
\end{align}

We apply the EM algorithm \cite{dempster1977maximum} to maximize the objective in the presence of latent variables. 
It is crucial to properly initialize the EM algorithm to avoid convergence to a local optima. 
The supervised identity linkage model is fully trained based on the ground truth set $\mb{A}$.  
The variance of positive samples ($\bs{\sigma_{p}}$) is initialized by  \(\bs{\sigma_{p}}^2 = \frac{1}{|\mb{A}|\cdot{d}} \sum_{i=1}^{|\mb{A}|}\|\bs{\mu_{si}} - \bs{\mu_{ti}}\|^2\), in which $d$ is the embedding dimension.  
Parameters $\bs{\hat{\mu}}$ and $\bs{\hat{\sigma}}$ are initialized by calculating the mean and variance of the entire target platform. 
 $\gamma$ is set to $0.5$. 
Algorithm \ref{alg:algorithm2} shows the training procedure. 
In $E$-step, we select a subset from the confident samples which are more likely to be generated from the positive distributions. 
Then in the  $M$-step, we use the selected subset to fine-tune the supervised linkage model and update the parameters in the two distributions. 
Through the iteratively training between $E$-step and $M$-step, we can learn better positive/noisy distributions and select more reliable candidates.  
From the results in Table $\ref{tab:examples_of_noises}$, with the noise filtering component (i.e., NSVUIL$_{sl}$), the accuracy of the confident pairs is significantly improved compared with NSVUIL$_{sup}$, which will facilitate the training in  next iterations.     


\begin{algorithm}[tb]
	\caption{Noise-aware self-learning part}
	\label{alg:algorithm2} 
	\begin{flushleft}
		\textbf{Require}: source network $S$, target network $T$, the annotation set $A$, the number of confident pairs $k_{c}$ and threshold  $\epsilon$ 
	\end{flushleft}
	\begin{algorithmic}[1] 
		\STATE Train a supervised linkage model based on Algorithm 1
		\STATE Select top $k_{c}$ unlabeled identity pairs with the highest linkage  probabilities along with their mean vectors $\{(\mu_{si}, \mu_{ti})\}$
		\STATE Parameters $\gamma, \sigma_{p}, \hat{\mu}, \hat{\sigma}$ initialization  
		\STATE $\gamma_{c}$ = 0.5, $\gamma_{p}$ = 0 \\ 
		\WHILE{$\left|\gamma_{c} - \gamma_{p}\right| > \epsilon$}
		\STATE \textit{// E-step}: \\
		\STATE  $g_i = p(q_i\! =\! 1 | \mu_{ti}, \mu_{si} ) = \frac{\gamma_{c}  \mathcal{N}(\mu_{si}, \sigma_{p}^2I)}{f(\mu_{ti}|\mu_{si})}$\\  
		\STATE $k_t= \sum_i 1 (g_i>0.5)$  \\
		
		\STATE \textit{// M-step}: \\
		\STATE $A_{c}$ = $\{i | g_i > 0.5 \}$ 
		\STATE Use $A$ and $A_{c}$ to fine-tune the supervised link model.\\
		\STATE $\sigma_{p}^2 = \frac{1}{d \cdot{k_t}}$ $\sum_{i|g_i > 0.5}\|\mu_{si} - \mu_{ti}\|^2$\\
		\STATE $\hat{\mu} = \frac{1}{(k_{c}-k_{t})} \sum_{i|g_{i} \leq 0.5} \mu_{ti}$\\
		$\hat{\sigma}^2 = \frac{1}{d{(k_{c}-k_{t})}} \sum_{i|g_{i} \leq 0.5} \|\hat{\mu} - \mu_{ti}\|^2$ \\
		\STATE $\gamma_{p} = \gamma_{c}$\\
		\STATE $\gamma_{c} = \frac{k_{t}}{k_{c}}$ \\
		\ENDWHILE
		\STATE Output $A_{c}$ as the filtered reliable samples.
	\end{algorithmic}
\end{algorithm}

\section{Experiments}
In this section, we evaluate the proposed NSVUIL model on multiple real-world social network datasets. 

\subsection{Datasets and Preprocessing} 

Our proposal is evaluated over five publicly available   datasets \cite{Li2019}, including two pairs of social networks and three pairs of academic co-author networks. 

Detailed statistics of the datasets are shown in Table \ref{tab:statistics_of_datasets}. 

\begin{itemize}
	\item \textbf{Twitter-Flickr}: 
	Twitter and Flickr are two popular social platforms. 
	In addition, about.me\footnote{\url{https://about.me}} website provides an intermediate platform for users to link and present their identities on different social websites, which can be viewed as the ground truth. 
	
	\item \textbf{Weibo-Douban}: 
	Sina Weibo  and Douban are two popular social platforms in China. Douban users can present their  Sina Webo accounts on the homepage, and thus this linkage information can be crawled as the ground truth. 
	
	\item \textbf{DBLP}: DBLP\footnote{\url{http://dblp.uni-trier.de/}}  is a computer science bibliography website. The dataset of DBLP is publicly available. 
	The published papers along with the authors in three years (2015, 2016, and 2017) are selected to form three co-author networks. 
	For each year, \emph{Yoshua Bengio} is selected as the center node, and then  a co-author subnetwork is constructed by locating the co-authors who can be reached within three steps from the center node.
	The published papers of one author in this year are viewed as his/her attributes. 
	Keywords in the papers published by a researcher are viewed as her demographic data. 
	The author identities from the DBLP dataset are considered as the ground truth.
\end{itemize}

These datasets include  social relations, demographic information and the posted microblogs.  
For the social network dataset, we randomly select 50 microblogs for each user. For the academic networks, the titles and abstracts of the published papers are viewed as microblogs.   
We utilize NTLK\footnote{https://www.nltk.org/} stemmer to process the crawled microblogs.

\begin{table}
	\centering
	\begin{threeparttable}
		\caption{Statistics of the datasets. }
		\begin{tabular}{P{1.6cm}P{2.2cm}P{2.2cm}P{0.7cm}}
			\toprule
			\multicolumn{1}{c}{Dataset}&\multicolumn{1}{c}{Source Network}&\multicolumn{1}{c}{Target Network}&\multicolumn{1}{c}{\#$_{Matched}$}\cr
			\midrule
			Twi.-Fli.&Twitter (3,259)&Flickr (4,308)&2,773\cr
			Wei.-Dou.&Weibo (4,119)&Douban (4,554)&3,235\cr
			DBLP15-16&DBLP15 (3,881)&DBLP16 (5,989)&1,852\cr
			DBLP16-17&DBLP16 (5,989)&DBLP17 (7,073)&2,570\cr
			DBLP15-17&DBLP15 (3,881)&DBLP17 (7,073)&1,492\cr
			\bottomrule
		\end{tabular}
		\label{tab:statistics_of_datasets}
		\vspace{-4mm}
	\end{threeparttable}
\end{table}   

\begin{table*}
	\centering
	\begin{threeparttable}
		\caption{User identity linkage performance with different $k$ ($Hit$-$Precision$ score).}
		\begin{tabular}{P{1.1cm}|P{0.6cm}P{0.6cm}P{0.6cm}|P{0.6cm}P{0.6cm}P{0.6cm}|P{0.6cm}P{0.6cm}P{0.6cm}|P{0.6cm}P{0.6cm}P{0.6cm}|P{0.6cm}P{0.6cm}P{0.6cm}}
			\toprule
			\multicolumn{1}{c}{}&\multicolumn{3}{c}{Twitter-Flickr}&\multicolumn{3}{c}{Weibo-Douban}&\multicolumn{3}{c}{DBLP15-16}&\multicolumn{3}{c}{DBLP16-17}&\multicolumn{3}{c}{DBLP15-17}\cr
			\midrule
			k&k=3&k=5&k=10&k=3&k=5&k=10&k=3&k=5&k=10&k=3&k=5&k=10&k=3&k=5&k=10\cr
			\midrule
			MAH&0.132&0.153&0.192&0.125&0.142&0.191&0.277&0.309&0.354&0.275&0.305&0.356&0.267&0.311&0.363\cr
			COSNET&0.144&0.187&0.236&0.132&0.161&0.194&0.292&0.330&0.373&0.288&0.332&0.386&0.289&0.338&0.375\cr
			IONE&0.161&0.196&0.242&0.150&0.189&0.232&0.302&0.347&0.397&0.308&0.345&0.396&0.310&0.352&0.377\cr
			CoLink&0.193&0.225&0.267&0.171&0.193&0.244&0.322&0.379&0.414&0.310&0.345&0.400&0.317&0.366&0.395\cr
			ULink&0.141&0.162&0.199&0.113&0.142&0.198&0.283&0.318&0.359&0.304&0.317&0.375&0.278&0.325&0.366\cr
			SNNA$_{u}$&0.228&0.244&0.295&0.215&0.246&0.282&0.342&0.388&0.437&0.323&0.353&0.427&0.331&0.376&0.423\cr
			SNNA$_{o}$&0.263&0.283&0.321&0.251&0.282&0.311&0.383&0.420&0.461&0.350&0.399&0.457&0.373&0.417&0.469\cr
			NSVUIL&\textbf{0.302}&\textbf{0.329}&\textbf{0.354}&\textbf{0.289}&\textbf{0.331}&\textbf{0.348}&\textbf{0.419}&\textbf{0.462}&\textbf{0.507}&\textbf{0.391}&\textbf{0.433}&\textbf{0.478}&\textbf{0.404}&\textbf{0.445}&\textbf{0.497}\cr
			\bottomrule
		\end{tabular}
		\label{tab:unsu}
		\vspace{-4mm}
	\end{threeparttable}
\end{table*}

\subsection{Baseline Methods}
We select several  state-of-the-art baselines including the semi-supervised and supervised models. 

\begin{itemize}
	
	\item \textbf{MAH} \cite{tan2014mapping} is a semi-supervised model that incorporates the network structure information and uses hypergraph to model the high-order relations. 
	
	\item \textbf{COSNET} \cite{zhang2015cosnet} is an energy-based model considering both local and global consistencies, which proposes an energy-based model to link user
	identities and develops an efficient subgradient algorithm to convert the original energy-based objective function into its dual form.
	
	\item \textbf{IONE} \cite{liu2016aligning}  adopts the representation learning approach to align users across multiple social networks, which solves
	both the network embedding problem and the user identity linkage problem under a unified
	optimization framework. 
	
	\item \textbf{CoLink} \cite{zhong2018colink} is a weakly-supervised model  employing a co-training algorithm to manipulate the attribute-based and relationship-based models, which reinforce each other iteratively in a co-training framework. 
	
	\item \textbf{ULink} \cite{mu2016user} introduces a new concept of ``Latent User Space” and optimizes objective function jointly with matched/unmatched pairs and intra-platform relation constraints across different platforms to conduct user identity linkage. 
	
	\item \textbf{SNNA$_{u}$} \cite{Li2019} aims to learn a projection function which can not only minimize the Wasserstein distance between the distributions of user identities in two social networks, but also incorporate the  annotations as the learning guidance. SNNA$_{u}$ is the unidirectional adversarial learning based alignment model.  
	
	\item \textbf{SNNA$_{o}$} \cite{mu2016user} is the extension of SNNA$_{u}$ with orthogonal restriction. 
\end{itemize}

\subsection{Parameter Setup} 
For the proposed NSVUIL model, we use the GLOVE word embeddings\footnote{\url{https://nlp.stanford.edu/projects/glove/}} pretrained from Twitter for English texts and use the word embeddings\footnote{\url{https://github.com/Embedding/Chinese-Word-Vectors}} for Chinese texts. The dimension of trainable vectors in the hierarchical user modeling part is set to 64. 
In the variational identity linkage module, the numbers of neural cells in the layers of encoder are  $\{64,32\}$, the weight $\lambda$ is set to 0.3, $\beta$ is set to 0.4. 
In the self-learning module, the number of high confident samples $k_{c}$ is set to $50$ and the threshold $\epsilon$ is set to 0.1.  
The iteration times of self-training is set to 3.  
For the CoLink model, we employ SVM  as the attribute-based model and randomly select the training seeds from the matched identity pairs.
The ULink model is trained by the  constrained concave convex procedure optimization.
For the SNNA models, the number of discriminator training times in an iteration $n_{d}$ is set to 5, the clipping parameter $c$ is set to 0.01, the annotation weigh $\lambda_{c}$ is set to 0.2 and the back-projection weight $\lambda_{r}$ is set to 0.3. 
Parameters in other baselines are set according to the original papers. 

\noindent \textbf{Evaluation Metric}
Following the previous work \cite{mu2016user},  $Hit$-$Precision$ is selected as the evaluation metric:
\begin{equation}
	h(x) = \frac{k-(hit(x) - 1)}{k}
\end{equation}
where $hit(x)$ is the rank position of the matched target user in the returned top-k candidate target identities. 
The $Hit$-$Precision$ is calculated by averaging scores of the matched identity pairs:  $\frac{\sum_{i=0}^{i=m}h(x_{i})}{m}$, in which $m$ is the number of source identities in the matched pairs.

\begin{table*}
	\centering
	\begin{threeparttable}
		\caption{User identity linkage performance with different training ratio $T_{r}$ ($Hit$-$Precision$ score).}
		\begin{tabular}{P{1.1cm}|P{0.7cm}P{0.7cm}P{0.8cm}|P{0.7cm}P{0.7cm}P{0.8cm}|P{0.7cm}P{0.7cm}P{0.8cm}|P{0.7cm}P{0.7cm}P{0.8cm}|P{0.7cm}P{0.7cm}P{0.8cm}}
			\toprule
			\multicolumn{1}{c}{}&\multicolumn{3}{c}{Twitter-Flickr}&\multicolumn{3}{c}{Weibo-Douban}&\multicolumn{3}{c}{DBLP15-16}&\multicolumn{3}{c}{DBLP16-17}&\multicolumn{3}{c}{DBLP15-17}\cr
			\midrule
			$T_{tr}$&$T_{tr}$=0.1&$T_{tr}$=0.3&$T_{tr}$=0.5&$T_{tr}$=0.1&$T_{tr}$=0.3&$T_{tr}$=0.5&$T_{tr}$=0.1&$T_{tr}$=0.3&$T_{tr}$=0.5&$T_{tr}$=0.1&$T_{tr}$=0.3&$T_{tr}$=0.5&$T_{tr}$=0.1&$T_{tr}$=0.3&$T_{tr}$=0.5\cr
			\midrule
			MAH&0.132&0.141&0.163&0.125&0.136&0.156&0.277&0.285&0.316&0.275&0.287&0.301&0.267&0.287&0.318\cr
			COSNET&0.144&0.155&0.179&0.132&0.143&0.166&0.292&0.325&0.344&0.288&0.297&0.314&0.289&0.306&0.328\cr
			IONE&0.161&0.174&0.197&0.150&0.168&0.183&0.302&0.324&0.351&0.308&0.323&0.369&0.310&0.346&0.363\cr
			CoLink&0.193&0.207&0.223&0.171&0.186&0.212&0.322&0.341&0.377&0.310&0.323&0.356&0.317&0.338&0.355\cr
			ULink&0.141&0.156&0.174&0.113&0.127&0.133&0.283&0.302&0.331&0.304&0.327&0.344&0.278&0.293&0.328\cr
			SNNA$_{u}$&0.228&0.233&0.256&0.215&0.231&0.247&0.344&0.366&0.390&0.323&0.341&0.375&0.331&0.344&0.371\cr
			SNNA$_{o}$&0.263&0.275&0.292&0.251&0.278&0.296&0.383&0.407&0.433&0.350&0.367&0.389&0.373&0.395&0.434\cr
			NSVUIL&\textbf{0.302}&\textbf{0.319}&\textbf{0.347}&\textbf{0.289}&\textbf{0.303}&\textbf{0.339}&\textbf{0.419}&\textbf{0.455}&\textbf{0.471}&\textbf{0.391}&\textbf{0.418}&\textbf{0.443}&\textbf{0.404}&\textbf{0.428}&\textbf{0.451}\cr
			\bottomrule
		\end{tabular}
		\label{tab:trainingratio}
	\end{threeparttable}
\vspace{-4mm}
\end{table*}

\subsection{Quantitative Evaluation}
For each dataset, $T_{tr}$ portion of aligned identity pairs are randomly selected as the training annotations, and $N_{te}$ linked pairs are  randomly selected as the test samples. 
Here $T_{tr}$ is fixed to 10\% and $N_{te}$ is set to 300. 
We repeat this process 5 times and report the average $Hit$-$Precision$ scores. 

Table \ref{tab:unsu} shows the experimental results with different settings of $k$. 
From the results, one can see that all models perform better on the co-author networks than the social networks, which may be due to the denser topological connections and comparatively formatted node attributes in the academic networks. 
CoLink model achieves the best performance among all sample-level based approaches (MAH, COSNET, IONE, CoLink and ULink) as it effectively captures the highly non-linear correlations between the node attributes and network topology.  
The adversarial learning-based methods (SNNA$_{u}$ and SNNA$_{o}$) achieve the best performance among all the baselines, which verifies the incorporation of distribution-level isomorphism contributes to better identity alignment.  
SNNA$_{o}$ outperforms SNNA$_{u}$ by around 4\%, which proves the effectiveness of orthogonal restriction. 
Our proposal consistently achieves the best performance  on all  datasets with different settings, and beats the best baseline SNNA$_{o}$ by more than 2\%. 
The improvement over the best performing baseline methods
(i.e., SNNA$_{o}$) is statistically significant (sign test, $p$-value $\leq$ 0.01).
Overall, the experimental results demonstrate the superiority of our proposal.  

We also conduct another experiment to study the performance of user identity linkage models given different training ratios $T_{tr}$. Parameter $k$ is fixed to 3. Training ratio $T_{tr}$ is increased from 0.1 to 0.5. Table \ref{tab:trainingratio} presents $Hit$-$Precision$ scores on five datasets. 
From the experimental results, one can clearly see that with the increase of training ratio, all models achieve better performance. It proves that annotations can provide task-relevant guidance, which is also the limitation of unsupervised approaches. 
The proposed NSVUIL model consistently outperforms baseline methods given different training ratios. 
The performance gap between NSVUIL and other models significantly increases given more annotations, which indicates the upper bound of NSVUIL is much higher.

\subsection{Ablation Study} 

Here we perform ablation study on the proposed NSVUIL model from different perspectives. Parameter $k$ in the $Hit$-$Precision$ measurement is fixed to 3, and the training ratio is set to 0.1. 
As the DBLP datasets share similar characteristics, here we only perform ablation studies over four datasets to save the spaces. 

\begin{table}
	\centering
	\begin{threeparttable}
		\caption{Ablation study on attention mechanism in the hierarchical user modeling part. }
		\begin{tabular}{P{1.5cm}P{1cm}P{1cm}P{1cm}P{1cm}}
			\toprule
			\multicolumn{1}{c}{Method}&\multicolumn{1}{c}{Twi.-Fli.}&\multicolumn{1}{c}{Wei.-Dou.}&\multicolumn{1}{c}{DBLP15-16}&\multicolumn{1}{c}{DBLP16-17}\cr
			\midrule
			BOW &0.183&0.178&0.223&0.205\cr
			Node2Vec &0.164&0.162&0.204&0.195\cr
			LSTM &0.203&0.207&0.257&0.240\cr
			TADW &0.224&0.237&0.284&0.269\cr
			\midrule
			NSVUIL$_{w/o\_ sa}$&0.284&0.272&0.361&0.364\cr
			NSVUIL$_{w/o\_ ma}$&0.284&0.274&0.369&0.353\cr
			NSVUIL$_{w/o\_ ua}$&0.275&0.266&0.355&0.375\cr
			NSVUIL$_{w/o\_ ca}$&0.278&0.267&0.352&0.378\cr
			\midrule
			NSVUIL&\textbf{0.302}&\textbf{0.289}&\textbf{0.419}&\textbf{0.391}\cr
			\bottomrule
		\end{tabular}
		\label{tab:usermodeling}
	\end{threeparttable}
\end{table}

\noindent \textbf{Hierarchical user modeling} 
Here we investigate the effectiveness of the hierarchical social user modeling part. 
Firstly, we aim to prove the usefulness of the attention mechanism at different levels.   
Comparison methods are defined as follows:
\begin{itemize}
	\item \textbf{BOW} is the bag-of-words model. The posted microblogs are represented as the bag of their words, disregarding grammar and word order but keeping multiplicity. 
	
	\item \textbf{Node2Vec} \cite{grover2016node2vec} is a network embedding approach to encode the topology into the low-dimensional embeddings. 
	
	\item \textbf{LSTM} \cite{hochreiter1997long} is a sequence encoding model. Here a two-layer LSTM model is employed to encode the semantic information. 
	
	\item \textbf{TADW} \cite{yang2015network} is a text-associated network embedding model to  capture the text and graph topology.  
	
	\item \textbf{NSVUIL$_{w/o\_}$} is the variation of our proposal without the attention mechanism in a specific layer ($sa$, $ma$, $ua$ and $ca$ denote the sentence modeling, microblog modeling, single user modeling and contextual user modeling layers, respectively). We use max-pooling instead of attention.     
\end{itemize} 

Table \ref{tab:usermodeling} presents the experimental results. 
The text-based modeling approaches (BOW and LSTM) outperform the pure topology method (Node2Vec), which demonstrates that semantic information plays a more significant role in linking identities than social relations. 
Without the attention strategy, the performance of NSVUIL drops remarkably. 
It verifies that attention strategy is critical to learn desirable user representations by selecting task-relevant informative entities (e.g., words, sentences and microblogs) from the noisy social network data. 
In addition, NSVUIL$_{w/o\_ua}$ and NSVUIL$_{w/o\_ca}$ perform worse than other variants, which means the high-level attentions are more important than the ones in low-levels.    

\begin{figure}
	\centering
	\includegraphics[width=0.40\textwidth]{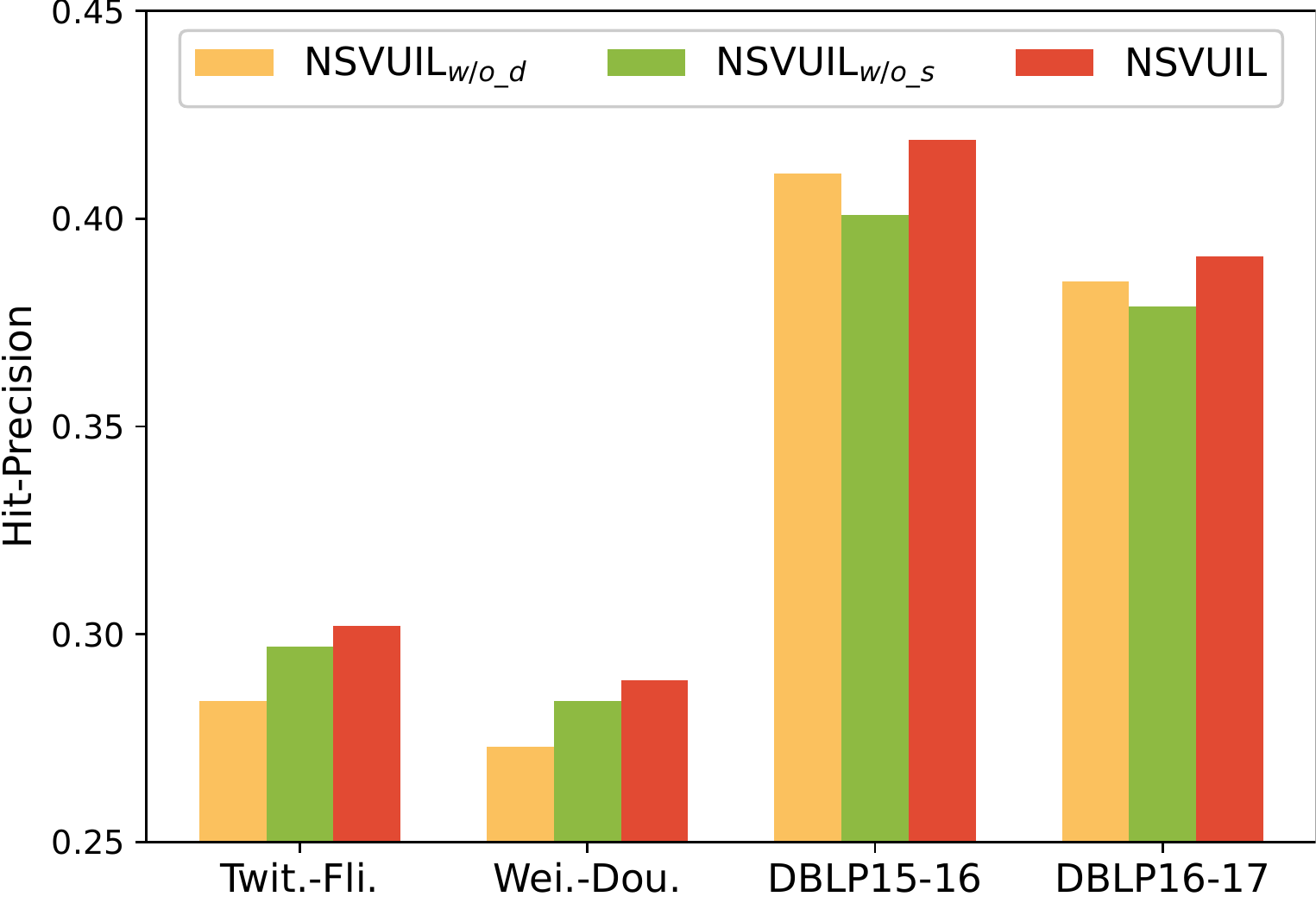}
	\caption{Ablation study on the categories of input data in the hierarchical user modeling part.  }
	\label{fig:userfeatures} 
	\vspace{-4mm}
\end{figure}

Table \ref{tab:usermodeling} proves the importance of attention schema, while the effectiveness of different types of input data is still obscure. 
Furthermore, we investigate the importance of social relations and demographic features.   
Two ablation models are proposed:
\begin{itemize}
	\item \textbf{NSVUIL$_{w/o\_d}$} is the variation of our proposal without the demographic features.  
	\item \textbf{NSVUIL$_{w/o\_s}$} is the variation of our proposal without the social relations. 
	Contextual user modeling layer is removed from Figure \ref{fig:usermodeling}.   
\end{itemize} 
Fig. \ref{fig:userfeatures} reports the experimental results. 
One can clearly see that the model performance consistently drops without social relations or the demographic features, which demonstrates these two types of data both contribute to capturing the cross-platform linkage signals. 
On the social network datasets, model performance presents a larger decline after removing the demographic features. 
This may be because demographic features denote the inherent properties of a natural person and are more likely to be shared across different platforms, while social relations are usually full of uncertainties and noises.   
Meanwhile, in the co-author networks, NSVUIL$_{w/o\_d}$ outperforms NSVUIL$_{w/o\_d}$, which shows that the co-author relations are more important than the demographic features. 
It is reasonable as the collaboration relationships among researchers are stronger and more trustable compared with the noisy social relations.

\begin{table}
	\centering
	\begin{threeparttable}
		\caption{Ablation study on the variational identity linkage module. }
		\begin{tabular}{P{1.5cm}P{1cm}P{1cm}P{1cm}P{1cm}}
			\toprule
			\multicolumn{1}{c}{Method}&\multicolumn{1}{c}{Twi.-Fli.}&\multicolumn{1}{c}{Wei.-Dou.}&\multicolumn{1}{c}{DBLP15-16}&\multicolumn{1}{c}{DBLP16-17}\cr
			\midrule
			SNNA$_{o}$&0.263&0.251&0.383&0.350\cr
			NSVUIL$_{d}$&0.284&0.271&0.396&0.378\cr
			NSVUIL$_{kl}$&0.286&0.276&0.403&0.387\cr
			NSVUIL&\textbf{0.302}&\textbf{0.289}&\textbf{0.419}&\textbf{0.391}\cr
			\bottomrule
		\end{tabular}
		\label{tab:uni}
	\end{threeparttable}
\end{table} 

\begin{figure}
	\centering
	\includegraphics[width=0.30\textwidth]{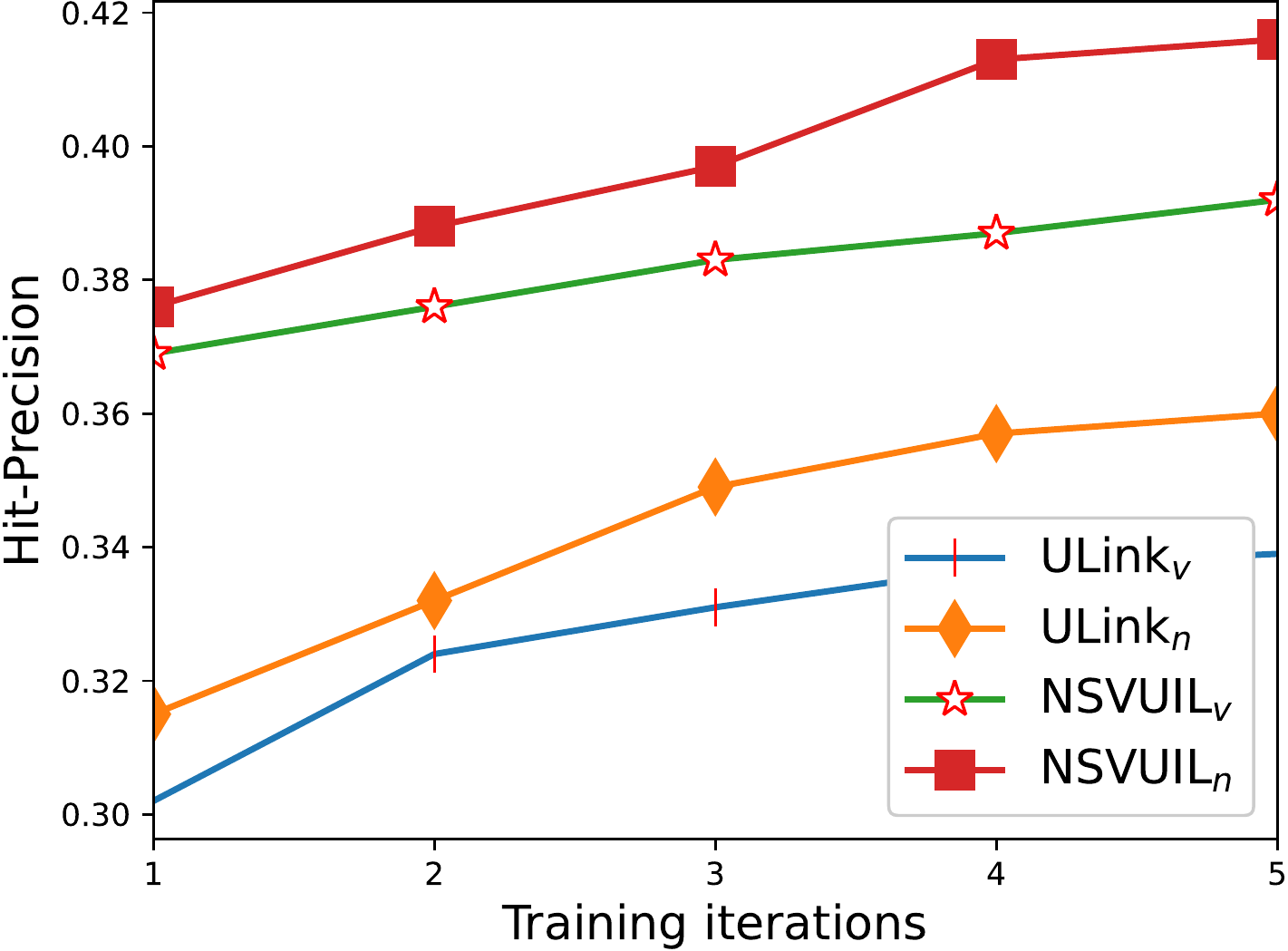}
	\caption{Training procedure of the self-learning module. }
	\label{fig:trainingiterations} 
	\vspace{-4mm}
\end{figure}

\begin{figure*}
	\centering
	\subfigure[Sensitivity of various methods on the sparsity of social relations.]{
		\label{fig:edgesparse} 
		\includegraphics[width=0.94\textwidth]{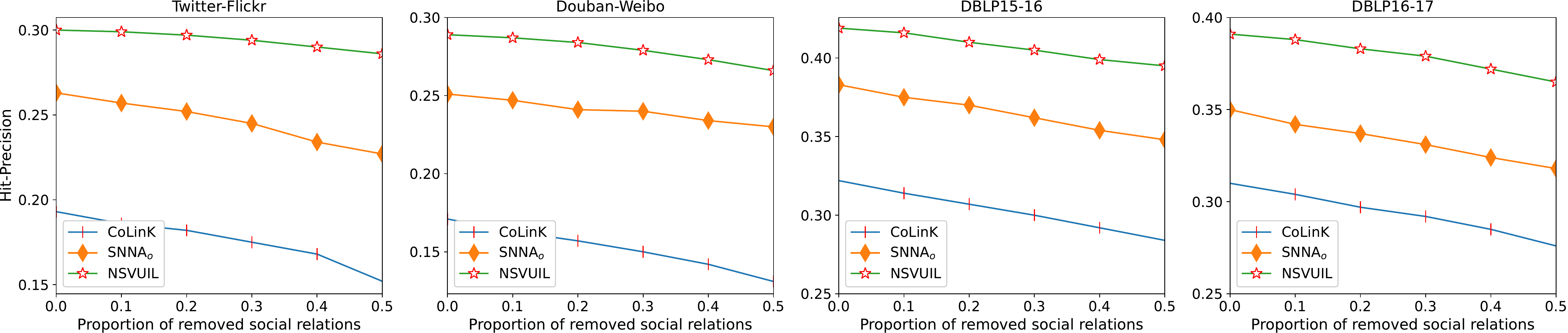}}
	\hspace{0.01in}
	\subfigure[Sensitivity of various methods on the sparsity of microblogs.]{
		\label{fig:textsparse} 
		\includegraphics[width=0.94\textwidth]{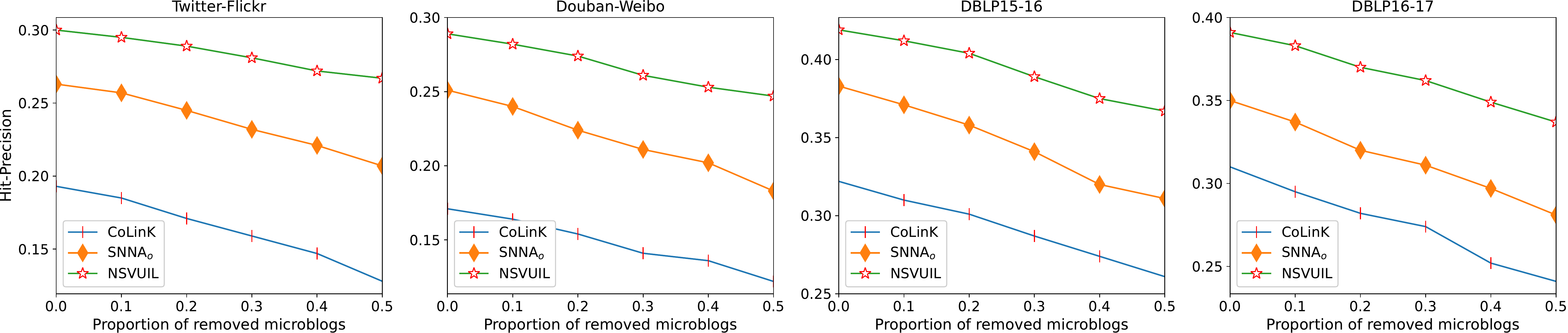}}
	\label{fig:sparse} 
	\caption{Identity linkage performance w.r.t the proportion of removed social relations and microblogs. }
	\vspace{-4mm}
\end{figure*}

\noindent \textbf{Variational identity linkage} 
Here we study the importance of the variational identity linkage module. 
Two ablation models are designed to evaluate the effectiveness of the proposed Wasserstein-based linkage loss: 
\begin{itemize}
	\item \textbf{NSVUIL$_d$} represents users as deterministic vectors (i.e., vector $\bm{z}$ in Figure \ref{fig:usermodeling}) instead of Gaussian distributions. Following previous works \cite{Li2019}, the L2-distance between the source embedding and target embedding is viewed as the linkage signal. The training objective function is also a negative-sample based loss similar to Formula (\ref{linkageloss}).   
	
	\item \textbf{NSVUIL$_{kl}$} utilizes the KL-divergence as the distribution measurement. We simply replace the $W_{2}$ function in Formula  (\ref{linkageloss}) by the KL-divergence calculation.    
\end{itemize} 
Table \ref{tab:uni} presents the experimental results of different ablation models. 
The strongest baseline SNNA$_{o}$ is also reported for comparison.  
One can see that the simplified version of our proposal NSVUIL$_d$ consistently outperforms SNNA$_{o}$ by nearly 4\%, which proves the effectiveness of  hierarchical user modeling and noise-aware self-learning modules. 
After representing users as Gaussian distributions, NSVUIL$_{kl}$ further improves the performance by a small margin. 
The improvements demonstrate the usefulness of Gaussian-based user representations, while the model performance is hindered by the inherent limitations of KL-divergence as KL-divergence is not symmetric and does not satisfy the triangle inequality \cite{zheng2019deep}.   
 NSVUIL  outperforms NSVUIL$_{kl}$ by nearly 2\% by enjoying the merits of W2 Wasserstein distance. 

\noindent \textbf{Noise-aware self-learning module}
In order to evaluate the effectiveness of the noise-aware learning module, we design the following four variations:  
\begin{itemize}
	\item \textbf{ULink$_{v}$}: ULink is the strongest supervised learning baseline. Here we combine ULink and vanilla self-learning strategy together as the ULink$_{v}$ model.     
	
	\item \textbf{ULink$_{n}$} is the combination of ULink and noise-aware self-learning. 
	
	\item \textbf{NSVUIL$_{v}$} replaces the noise-ware learning part in NSVUIL model with the vanilla self-learning module. 
	
	\item \textbf{NSVUIL$_{n}$} is the proposed NSVUIL model.  
	
\end{itemize} 
Evaluation is conducted on the DBLP15-16 dataset, $k$ is set to $3$ and $T_{tr} = 0.1$. We process the self-learning module 5 times. After each self-learning process is finished, we record the Hit-Precision scores of the corresponding checkpoint. 
Results are shown in Fig. \ref{fig:trainingiterations}. 
One can see that noise-aware self-learning outperforms vanilla self-learning with both supervised learning models, which proves our proposal can effectively remove the noises from the confident pairs and thus contribute to improving the linkage performance. 
Given the same self-learning settings, the proposed NSVUIL model consistently outperforms ULink model, which further demonstrates the proposed attention-based modeling part can learn high quality and task-relevant user representations.    
Another interesting observation is the performance increasing trend of the vanilla self-learning module slows down from the second 
iteration, while the noise-aware one keeps an almost similar growth trend.
This may be because early mistakes affect the learning of supervised linkage model in the following iterations and further lead to the failure of the self-learning.

\subsection{Quantitative Evaluation with Sparse Data} 
As discussed in the introduction section, social data tend to be sparse and uncertain. 
Thus, we will evaluate the robustness of NSVUIL in the scenario of sparse data in this subsection. 

\noindent \textbf{Evaluation with sparse social relations.} 	
We randomly remove $R_{r}$ percentage of social relations from the original social networks and then perform social identity linkage. 
Parameter $k$ is set to $3$ and $T_{tr} = 0.1$.  
We increase $R_{r}$ from 10\% to 50\% to evaluate the performance of various methods under different proportions of the removed relations. 

Figure \ref{fig:edgesparse} presents the results. 
One can see that with the increase of the removed social relations, the performance of all methods drops. 
It demonstrates that social relations are important to the successful linkage. 
The proposed NSVUIL model consistently outperforms other methods and it presents the least performance decline on all datasets. 
By representing the users as Gaussian distributions instead of deterministic vectors, NSVUIL owns powerful expressive capacity and tends to be more robust, which ensures its stable performance. 
Furthermore, the noise-aware self-learning mechanism precisely augments the linkage signals in the unsupervised learning manner, which also contributes to alleviating the reliance of model performance on the quality of the input data. 
The performance drop on co-author networks is more significant than the ones on social networks, which demonstrates that the co-author relations are more stable and trustable compared with the social relationships.  

\begin{figure}
	\centering
	\subfigure[Reconstruction weight $\beta$.]{
		\label{fig:para1} 
		\includegraphics[width=0.23\textwidth]{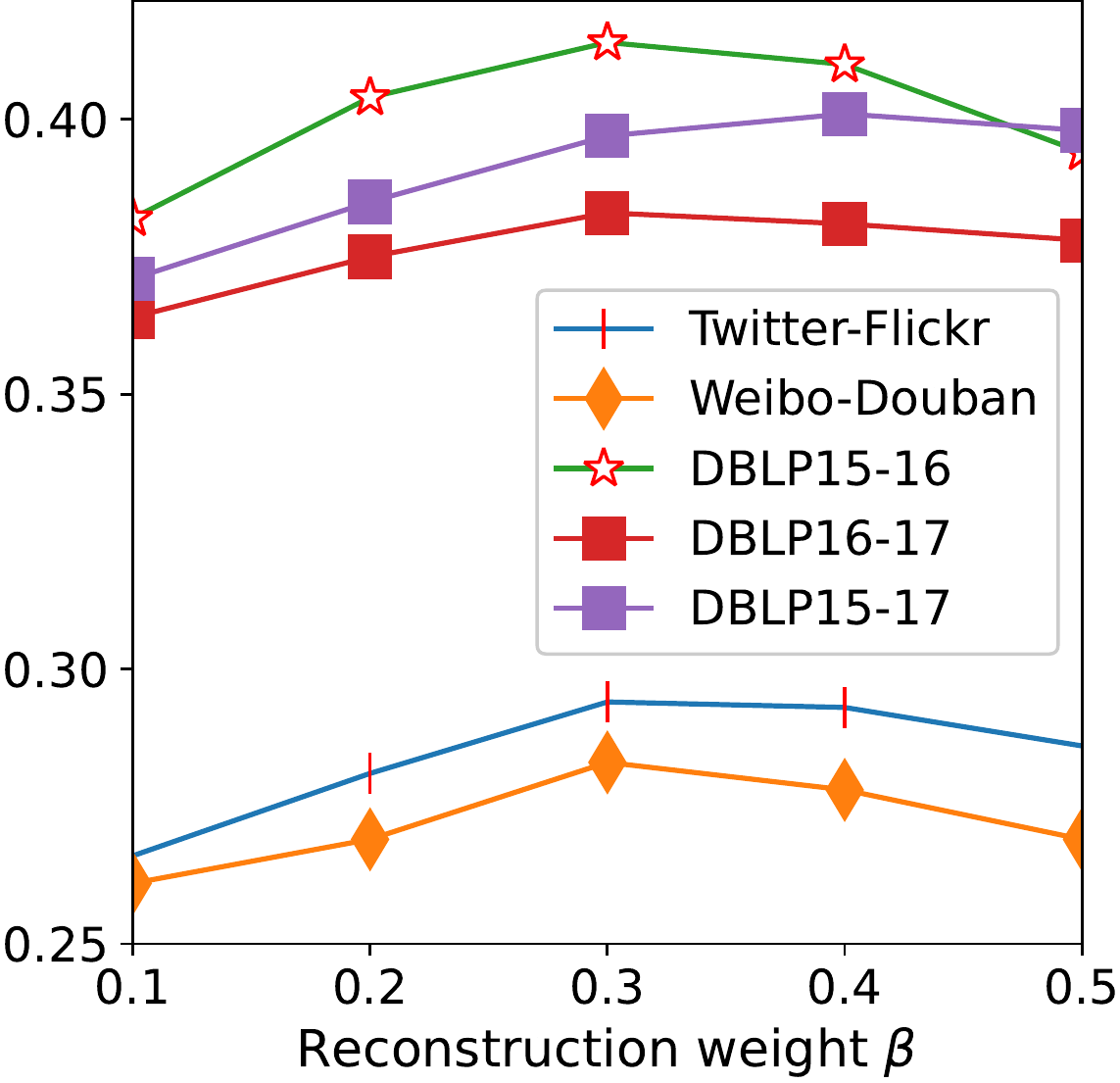}}
	\hspace{0.01in}
	\subfigure[Number of confident samples $k_{c}$.]{
		\label{fig:para2} 
		\includegraphics[width=0.23\textwidth]{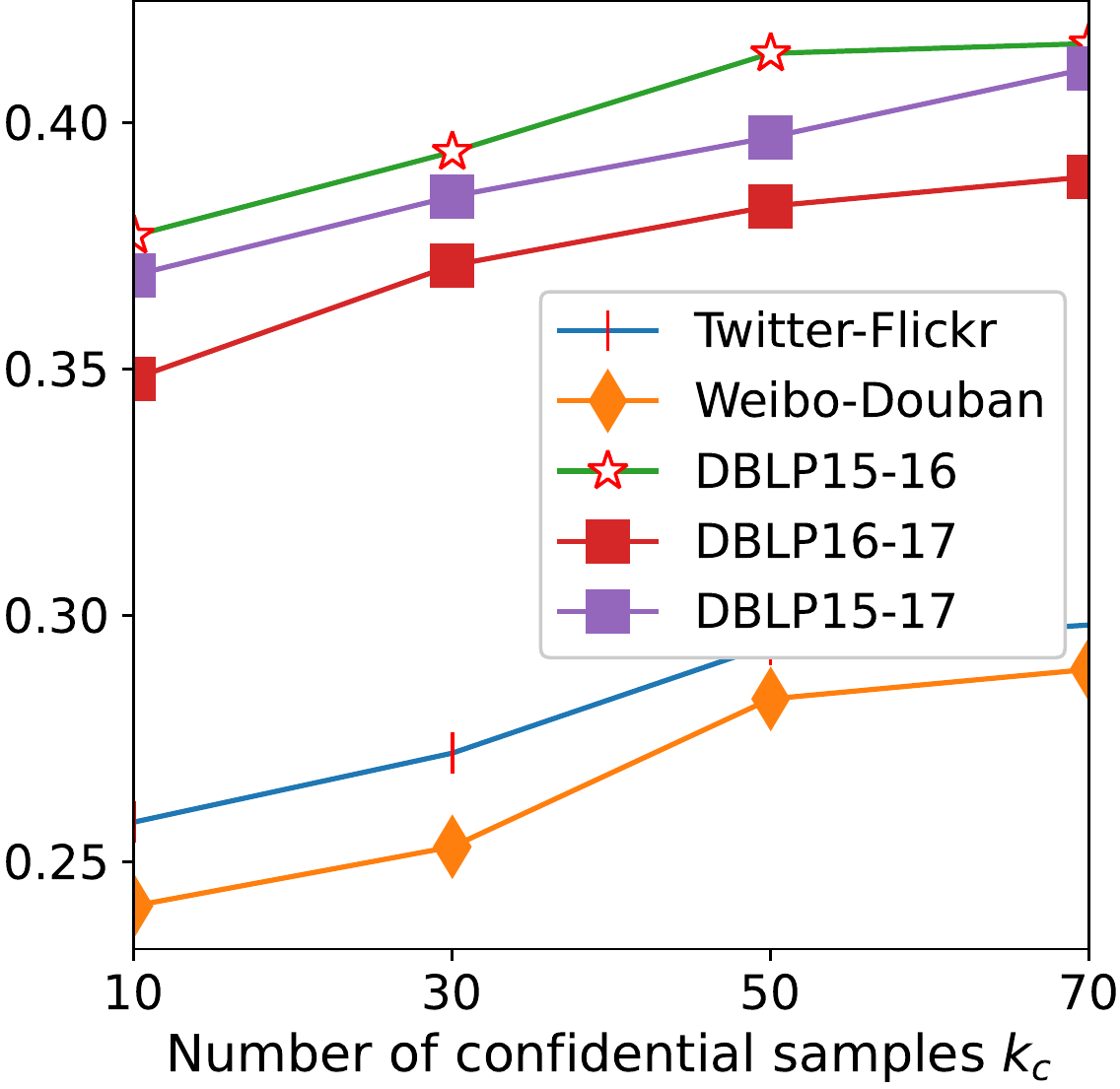}}
	\caption{Parameter sensitivity analysis. }
	\label{fig:parameteranalysi} 
	\vspace{-4mm}
\end{figure}

\noindent \textbf{Evaluation with sparse tweets.} 
Similarly as above, we randomly remove $R_{t}$ (from 10\% to 50\%) percentage of microblogs from the original posted texts, and report the performance of different methods on the remained sparse data.
Figure  \ref{fig:textsparse} reports the results. 
One can see that the removed microblogs have a much larger impact on the linkage performance than the social relations, which demonstrates the text information contributes more to the identity linkage than the social relationships. 
Although NSVUIL presents a performance drop trend, it still consistently outperforms other baselines. 
With the increase of $R_{t}$, the performance gap between the proposed NSVUIL model and the strongest baseline SNNA$_{o}$ is enlarged from 4\% to 7\%, which verifies NSVUIL is robust to the sparse data.

\subsection{Parameter Sensitivity Study} 
Here we study the performance sensitivity of NSVUIL model on two core parameters: the reconstruction weight $\beta$ and the number of confident samples $k_{c}$. 
Training ratio $T_{tr}$ is set to 10\% and  $k$ is fixed as 3. 
$\beta$ varies from 0.1 to 0.5, and $k_{c}$ is set from 10 to 70. Hit-Precision scores under different settings on five datasets are recorded. 
Fig. \ref{fig:parameteranalysi} presents the experimental results.
From the left sub-figure, one can see that with the increase of $\beta$, the performance over all the datasets first increases and then decreases, which demonstrates appropriate reconstruction constraint may benefit the alignment performance. 
However,  a larger $\beta$ will lead the training procedure to focus on the reconstruction task, which may interrupt and slow down the optimization to reach the  supervised linkage objective. 
From the right sub-figure, one can see that with the increase of $k_{c}$, the performance over all the datasets first increases and then keeps steady or slightly increases. 
It means that the enlarging of  $k_{c}$ contributes to better alignment performance at the beginning as more quality candidates are provided.  
A larger $k_{c}$ also leads to more time consumption, and thus an appropriate $k_c$ needs to be carefully chosen to balance the model efficiency and effectiveness.

\begin{figure}
	\centering
	\includegraphics[width=0.42\textwidth]{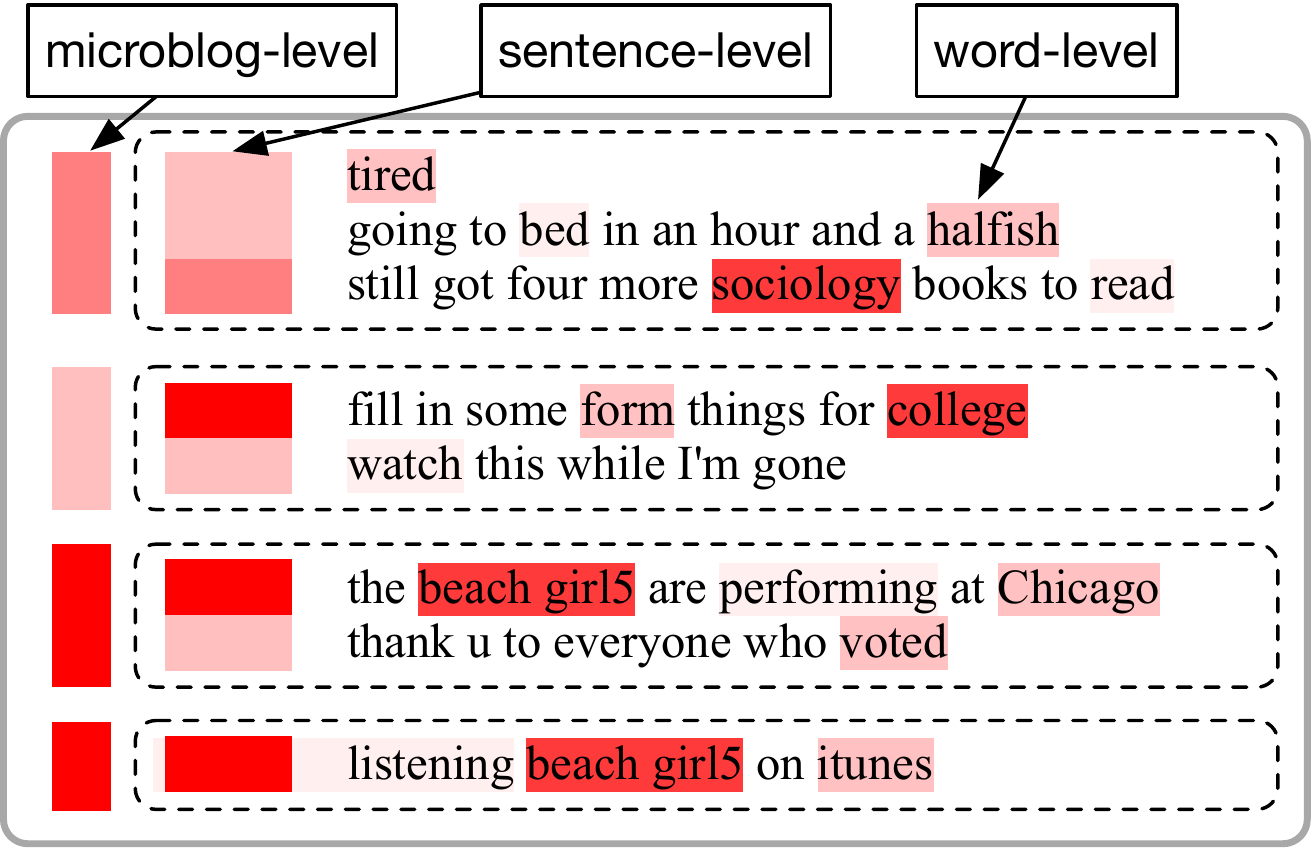}
	\caption{The visualization of attention scores in different layers. }
	\label{fig:visual} 
	\vspace{-4mm}
\end{figure}

\subsection{Efficiency Analysis} 
In this subsection we analyze the efficiency of the proposed NSVUIL model. 
The strongest baselines (SNNA$_{u}$ and SNNA$_{o}$) are selected as the comparison methods. 
The iteration times of self-training is set to 3.  
Running time reported in this subsection is performed on a Ubuntu 64-Bit Linux workstation with 4-core Intel Core(TM) i7-6700 2.40 GHz, 128 GB memory and two NVIDIA Tesla P100 GPUs. Two social network pairs and two academic network pairs are selected as the evaluation datasets. 
Table \ref{tab:efficency} shows the running time of different models. 
There is a explicit linear relationship between the running time and the size of dataset. 
SNNA$_{o}$ is slower that SNNA$_{u}$ model as it adds extra computations to ensure the learned projection matrix to be orthogonal. 
One can see that the NSVUIL is faster than the SNNA models. 
This is reasonable as SNNAs  are adversarial learning based models, which are time-consuming to seek a Nash equilibrium through the competition between the generator and discriminator. 
Overall, NSVUIL model outperforms strongest baselines with less training time, which proves its superiority.

\begin{table}
	\centering
	\begin{threeparttable}
		\caption{Training time (minutes) of different methods. }
		\begin{tabular}{P{1cm}P{1cm}P{1cm}P{1cm}P{1cm}}
			\toprule
			\multicolumn{1}{c}{Method}&\multicolumn{1}{c}{Twi.-Fli.}&\multicolumn{1}{c}{Wei.-Dou.}&\multicolumn{1}{c}{DBLP15-16}&\multicolumn{1}{c}{DBLP16-17}\cr
			\midrule
			SNNA$_{u}$&17.72&19.75&19.13&21.32\cr
			SNNA$_{o}$&19.63&21.24&20.38&23.57\cr
			NSVUIL&\textbf{11.34}&\textbf{14.13}&\textbf{12.56}&\textbf{16.87}\cr
			\bottomrule
		\end{tabular}
		\label{tab:efficency}
		\vspace{-4mm}
	\end{threeparttable}
\end{table} 

\subsection{Visualization}

In this subsection, we conduct a case study to further explore whether our approach can select important words, sentences and microblogs to learn informative user representations for social identity linkage. 
The visualization of the attention weights in the word-, sentence- and microblog-level attention networks are shown in Fig. \ref{fig:visual}. 
Each line indicates one sentence and the text inside a dotted box forms a microblog. 
Four microblogs are published by the same Twitter user. 
Darker color represents higher attention weights.
One can see that the attention networks can effectively select and attend to important entities at different levels. 
Based on the highlight areas, we can infer that the user is a college student who is fond of Beach Girl5 and sociology.
With this  explicit user figure, the linkage model can  easily match her to the correct Flickr user.

\section{Conclusion}
In this paper, we study the task of semi-supervised user identity linkage. 
In order to alleviate the challenge of annotation scarcity, a novel self-learning based approach is proposed to leverage the unlabeled data and few annotations. 
Specifically, a hierarchical attention-based user modeling module is proposed to effectively capture the semantic and social relation information. 
Then, a variational linkage model is designed to match identities according to the ground truth. 
Each social identity is represented by a Gaussian distribution to capture the uncertainty in the social networks. 
Finally, we employ a noise-aware self-learning strategy to improve the quality of pseudo annotations in the manner of generative latent variable based learning.       
Our proposal is extensively evaluated over five real-life datasets. 
Experimental results demonstrate that NSVUIL model consistently achieves the best linkage performance.

\bibliographystyle{IEEEtran}
\bibliography{bibfile}

\end{document}